\documentclass[aip,jcp,reprint]{revtex4-1}
\usepackage{yhmath}
\usepackage{graphicx}
\usepackage{color}
\usepackage{booktabs}
\usepackage[all]{xy}
\usepackage{latexsym}
\usepackage{xfrac}
\usepackage{calc}
\usepackage{rotating}
\usepackage[version=4]{mhchem}
\usepackage{siunitx}
\usepackage{amsmath,amsxtra,amsfonts,amssymb,amscd}
\usepackage{stackrel}
\usepackage{bbm}
\usepackage{bm}
\usepackage{enumerate}
\usepackage[cal=pxtx]{mathalfa}
\usepackage{array,hhline}
\newcolumntype{M}[1]{>{\centering\arraybackslash}m{#1}}
\usepackage{paralist}
\usepackage{accents}
\usepackage{epsf}
\usepackage{tikz}
\usepackage{pgfplots}
\pgfplotsset{width=6cm,compat=newest}
\DeclareMathAccent{\wtilde}{\mathord}{largesymbols}{"65}
\definecolor{lightblue}{rgb}{0.65,0.80,0.89}
\definecolor{cyan}{rgb}{0.,1.,1.}
\definecolor{softblue}{rgb}{0.21,0.49,0.72}
\definecolor{bordeaux}{rgb} {0.72,0.21,0.49}
\definecolor{indigo}{rgb}{0.45,0.13,0.74}
\definecolor{softgreen}{rgb}{0.3,0.67,0.29}
\definecolor{maroon}{rgb}{0.4,0.027,0.28}
\definecolor{magenta}{rgb}{1,0,1}

\def\d {{\rm d}}

\def\d {{\rm d}}
\def\thetak{{\vartheta}}
\def\phik{{\varphi}}

\def\utilde#1{\underaccent{\wtilde}{#1}}
\def\u2tilde#1{\underaccent{\sim\!\!\sim}{#1}}
\def\uttilde#1{\underaccent{\wtilde}{\underaccent{\wtilde}{#1}}}
\def\basis{{\mathfrak{B}}}

\def\tttt#1{\uttilde{\mathcal{#1}}}
\def\uvec#1{{\hat{\underline{#1}}}}

\DeclareMathOperator{\trace}{Tr}
\DeclareMathOperator{\Iso}{Iso}

\renewcommand{\tensor}[1]{{\underline{\underline{#1}}}}
\renewcommand{\vec}[1]{{\underline{#1}}}

\begin{document}

\title{Stress correlations in glasses}
\author{Ana\"el Lema\^{\i}tre}

\affiliation{Laboratoire Navier, UMR 8205, \'Ecole des Ponts, IFSTTAR, CNRS, UPE, Champs-sur-Marne, France}

\date{\today}

\begin{abstract}
 We rigorously establish that, in disordered three-dimensional (3D) isotropic solids, the stress autocorrelation function presents anisotropic terms that decay as $1/r^3$ at long-range, with $r$ the distance, as soon as either pressure or shear stress fluctuations are normal. By normal, we mean that the fluctuations of stress, as averaged over spherical domains, decay as the inverse domain volume. Since this property is required for macroscopic stress to be self-averaging, it is expected to hold generically in all glasses and we thus conclude that the presence of $1/r^3$ stress correlation tails is the rule in these systems. Our proof follows from the observation that, in an infinite medium, when both material isotropy and mechanical balance hold, (i) the stress autocorrelation matrix is completely fixed by just two radial functions: the pressure autocorrelation and the trace of the autocorrelation of stress deviators; furthermore, these two functions (ii) fix the decay of the fluctuations of sphere-averaged pressure and deviatoric stresses for windows of increasing volume. Our conclusion is reached because, due to the precise analytic relation (i) fixed by isotropy and mechanical balance, the constraints arising via (ii) from the normality of stress fluctuations demand the spatially anisotropic stress correlation terms to decay as $1/r^3$ at long-range. For the sake of generality, we also examine situations when stress fluctuations are not normal.
\end{abstract}

\maketitle

\section{Introduction}

The local stress field of inherent states (or ISs) produced from supercooled liquids was recently found, both in dimension $d=2$~\cite{Lemaitre2014} and $d=3$~\cite{Lemaitre2015}, to present anisotropic spatial correlations that decay as $1/r^d$ at long range, with $r$ the distance. These observations were obtained in studies examining the role of elasticity in supercooled relaxation, a process involving local rearrangements that leave long-range stress imprints in the surrounding medium~\cite{DyreOlsenChristensen1996,Dyre1999,ChattorajLemaitre2013,JensenWeitzSpaepen2014,Lemaitre2014,Lemaitre2015,IllingFritschiHajnalKlixKeimFuchs2016}. In this context, the observed long-range stress correlations were interpreted as a resulting from the accumulation of rearrangements and their elastic strains in the studied supercooled conditions~\cite{Lemaitre2014,Lemaitre2015}.


It is crucial to assess whether long-range stress correlations are specific to certain types of glasses, such as those previously studied (obtained after rapid quenches from supercooled conditions, using Lennard-Jones systems, which are nearly incompressible), or instead should be found in disordered isotropic solids under broad conditions. This is important because the presence of power law correlations implies that distant regions of space are not independent, which calls into question the well-definedness of the thermodynamic limit, and may have wide-ranging consequences. For example, the long-range character of stress correlations in glasses was suggested~\cite{GelinTanakaLemaitre2016} to cause an excess of sound attenuation compared with the Rayleigh prediction~\cite{MarruzzoSchirmacherFratalocchiRuocco2013}, a finding which may also have implications on thermal transport. However, these long-range correlations have never been probed in the specific materials such as silica or glycerol that show this attenuation excess~\cite{MonacoGiordano2009,BaldiGiordanoMonaco2011,RutaBaldiScarponiFiorettoGiordanoMonaco2012}; their existence is these systems, hence, remains to be established.

The existence of long-ranged stress correlations is also susceptible to raise a variety of issues in the context of supercooled liquids. The latter, indeed, are known to remain at almost all times in the vicinity of ISs, and to evolve via thermally activated hopping events. The relaxation of the parent stress, which ultimately determines liquid viscosity~\cite{FurukawaTanaka2009b,LevashovMorrisEgami2013,WuIwashitaEgami2015,Levashov2014,Levashov2017}, therefore tracks that of the IS stress at long times. Moreover, since local IS stress inevitably biases the activation barriers, it is susceptible to affect the dynamic relaxation processes, and the question arises whether its long-range correlated nature may play a role in cooperativity.

In a recent study~\cite{Lemaitre2017}, we showed that, in two dimensions, long-range stress correlations are an analytical consequence of the conjunction of three expected properties of glassy systems: mechanical balance, material isotropy, and the normal decay of pressure fluctuations. By this third condition we mean that the pressure averaged over circular observation windows in an infinite medium presents fluctuations that decay as the inverse window area. Our argument explained that similar long-range correlations had been found in 2D granular materials~\cite{HenkesChakraborty2009} near the jamming point~\cite{LiuNagel1998}. It was not initially apparent that the stress correlations observed near jamming and in supercooled ISs had a common origin, since Ref.~\cite{HenkesChakraborty2009}, was motivated as a test of Edwards' theory~\cite{EdwardsOakeshott1989}, a heuristic approach at constructing a statistical physics framework for rigid particles at the jamming point, and viewed these long-range correlations as evidence for the specific assumptions introduced in this framework. Our 2D study demonstrated that the long-range correlations are a very general property and do not constitute any evidence for Edwards' theory.

The question remains whether long-range stress correlations exist in 3D glasses (and granular systems) under the same general conditions. Here, we will show that they do, yet with one important distinctive feature. Specifically, we will show that, as in 2D, material isotropy and mechanical balance tightly constrain the overall stress autocorrelation matrix in three dimensions. In 2D, these constraints cause the whole stress autocorrelation function (a field of a priori six components) to be fixed by the pressure autocorrelation alone---which is a radially symmetric scalar function~\cite{Lemaitre2017}. As a consequence, pressure and deviatoric stress fluctuations are in a fixed ratio. In 3D, in contrast, under these same conditions of mechanical balance and material isotropy, the 3D pressure autocorrelation $C_0$ and the (normalized) trace of the autocorrelation of deviators $C_0'$, which are both radially symmetric, remain independent from each other, which entails that the pressure and deviatoric stress fluctuations are decoupled. Nevertheless, these two radial scalar functions, together, fix the rest of the autocorrelation matrix (a field of a priori 21 components).

Our presentation will emphasize that the constraints brought by material isotropy and mechanical balance leave a certain leeway concerning the spatial decays of the stress autocorrelation at long range. But we will prove that as soon as the fluctuations of either the sphere-averaged pressure or the sphere-averaged deviatoric stresses are normal---i.e. decay as the inverse observation volume---then the anisotropic part of the 3D stress autocorrelation decays as $1/r^3$ at long-range in space. This holds in particular when $C_0$ and $C_0'$ are short-ranged, a rather remarkable and counterintuitive fact.

In this regard, let us observe that, although the two most often cited defining properties of glasses are mechanical stability (they are solids), and the absence of any long-range crystalline, quasi-crystalline, or orientational order, which guarantees material isotropy, structural disorder is also widely expected to cause the fluctuations of certain window-averaged quantities such as density, energy, or stress, to decay as the inverse volume of the observation window. The only physical assumption that stress fluctuations are normal in glasses then suffices to conclude that the existence of $1/r^3$ anisotropic correlation tails is the rule for these systems.

It should be noted that, the fluctuations of window-averaged quantities do not decay as the inverse volume of the observation window in all isotropic systems. They may decay more slowly, e.g. near critical points, or in as-quenched systems from fully random states~\cite{WuKarimiMaloneyTeitel2017}, or may decay faster than the inverse window volume. This latter behavior is observed for density fluctuations (it is then called hyperuniformity) in certain exotic systems, which are, in some cases, random and isotropic~\cite{Torquato2018}. We are not aware of any system which is isotropic and exhibits a hyperuniform stress field, but will consider their existence as a possibility for the sake of generality. Our analysis handles straightforwardly the two hypothethical cases when the window-averaged stress fluctuations scale anomalously. We will thus show that the $1/r^3$ anisotropic tails are found in isotropic solids \emph{only} when the fluctuations of window-averaged stresses present the normal inverse-volume decay. In that sense, not only does our work demonstrate that the presence of $1/r^3$ power-law correlations is the rule in  glasses, but also that these tails are the signature of normal stress fluctuations in isotropic solids.\\

Our analysis will largely rely upon a formalism we constructed previously~\cite{Lemaitre2015} to deal with the transformation of second and fourth order tensors under rotations, and which helped us specify how material isotropy constrains the structure of the stress autocorrelation tensor. Since this formalism is unfamiliar and instrumental in key steps of our present argument, we need to recall its main concepts in Sec.~\ref{sec:formalism}. Meanwhile, we will introduce improvements in our terminology and notation. We also find useful to recall, for the sake of completeness, the rigorous derivation from Ref.~\cite{Lemaitre2015} concerning the consequences of material isotropy on stress correlations. It will be presented at the beginning of Sec.~\ref{sec:materialisotropy}  and will help up proceed in Section~\ref{sec:isotropy} to the identification and analysis of the isotropic part of the stress autocorrelation tensor. In Section~\ref{sec:thermo} we show how this isotropic part, which only involves the two functions $C_0$ and $C_0'$, is related to the fluctuations of the window-averaged stress; we will show in particular that the decay of the averaged stress fluctuations is normal iff the (3D) Fourier transforms of these functions, $\widehat{C}_0(k)$ and $\widehat{C}_0'(k)$, are continuous (and hence finite) at $k=0$. The core results of the paper are presented in Section~\ref{sec:compounding}, where we first show that when both isotropy and mechanical balance hold, the functions $\widehat{C}_0(k)$ and $\widehat{C}_0'(k)$ fix the complete stress autocorrelation matrix; and then systematically analyze how the spatial decay of the correlations in real-space is set by the low-$k$ behavior of these functions.

\section{Formalism}
\label{sec:formalism}


\subsubsection{Stress and its autocorrelation matrix}

The basic idea of our approach~\cite{Lemaitre2015} is to use a vector representation for stress that will later permit us to express rotations using matrix-vector products. It will prove tremendously useful to use vector components that, like spherical harmonics, correspond to eigenspaces of axial rotation around the $z$ axis~\cite{Rose1957}. More precisely, the stress components we choose are real-valued, and hence analogous to tesseral harmonics, or equivalently Stevens operators~\cite{Stevens1952,EgamiSrolovitz1982}.

In a reference Cartesian basis, we define the \emph{tesseral components}\footnote{We previously said \emph{spherical}, but the word \emph{tesseral} is more appropriate since they are real-valued.} of an arbitrary stress tensor $\tensor\sigma$ as:
\begin{equation}
\label{eq:stress}
\begin{split}
\sigma_1&=-\frac{1}{\sqrt{3}}\,\left(\sigma_{xx}+\sigma_{yy}+\sigma_{zz}\right)\\
\sigma_2&=-\frac{1}{\sqrt{6}}\,\left(\sigma_{xx}+\sigma_{yy}-2\,\sigma_{zz}\right)\\
\sigma_3&=\sqrt{2}\,\sigma_{yz}\\
\sigma_4&=\sqrt{2}\,\sigma_{xz}\\
\sigma_5&=\sqrt{2}\,\sigma_{xy}\\
\sigma_6&=\frac{1}{\sqrt{2}}\,\left(\sigma_{xx}-\sigma_{yy}\right)\\
\end{split}
\end{equation}
The set of these components defines a so-called \emph{tesseral vector} denoted $\utilde\sigma=(\sigma_1,\ldots,\sigma_6)$.\\

Stress autocorrelation is the matrix:
\begin{equation}
  \label{def:C}
  \uttilde{C}(\vec r)\equiv\langle\utilde\sigma(\vec r_0+\vec r;t)\,\utilde\sigma(\vec r_0;t)\rangle_c
\end{equation}
with $\langle AB \rangle_c=\langle AB\rangle-\langle A\rangle\langle B \rangle$ the second cumulant. We use juxtaposition to denote the tensor product, so that the above expression means that the matrix components of $\uttilde{C}$ are $C_{ab}=\langle\sigma_a(\vec r_0+\vec r;t)\,\sigma_b(\vec r_0;t)\rangle_c$. Let us recall that a general rank-2 tensor has 9 components, and a general fourth order tensor 81. Stress has 6 components because it is symmetric. Therefore, the stress autocorrelation is a $6\times6$ matrix and has a priori 36 components.

We focus in this paper on the case of systems that are invariant by translation and spatial inversion. The former property entails that $\uttilde{C}$ does not depend on $\vec r_0$, as our notation already suggests, and also that $C_{ab}(\vec r)=C_{ba}(-\vec r)$. Spatial inversion symmetry adds that $C_{ab}(\vec r)=C_{ab}(-\vec r)$ which, compounded with the previous relation, guarantees matrix symmetry, $C_{ab}=C_{ba}$. 

\subsubsection{Rotations}

We work with ``passive'' rotations, i.e. changes of basis vectors, the system remaining fixed.
The reference basis of the lab frame is denoted $\basis=(\vec e_x,\vec e_y,\vec e_z)$. To a rotation matrix $\tensor R$ we associate the rotated basis, $\basis^{\tensor R}=(\tensor R^T.\vec e_x,\tensor R^T.\vec e_y,\tensor R^T.\vec e_z)$ with $^T$ the transpose. With this convention, a given material point of coordinates $\vec r$ in basis $\basis$ has the coordinates $\vec r'=\tensor R\cdot\vec r$ in $\basis^{\tensor R}$. Similarly, a stress tensor of matrix form $\tensor\sigma$ in $\basis$ becomes:
\begin{equation}
\tensor\sigma'=\tensor R\cdot\tensor\sigma\cdot\tensor R^T
\end{equation}
in $\basis^{\tensor R}$. The $\tensor\sigma\to\tensor\sigma'$ operation is a linear transformation: using our vector form of stress is hence writes it as a matrix-vector product:
\begin{equation}
\utilde\sigma'=\tttt{D}\cdot\utilde\sigma
\end{equation}
To understand what is the structure of $\tttt{D}$ (we will not write its explicit form), let us first recall that any rotation $\tensor R$ can be decomposed into a series of three axial rotations. In the so-called ZYZ decomposition, these axial rotations are: i) a rotation $\tensor R^z(\phi)$ about $z$ by $\phi$; ii) a rotation about the new Y-axis, say $y'$, by $\theta$; iii) a rotation about the new Z-axis, say $z''$, by $\psi$, so that:
\begin{equation}
\tensor R=\tensor R^{z''}(\psi)\cdot\tensor R^{y'}(\theta)\cdot\tensor R^z(\phi)\\
\end{equation}
It is a classical result that this is equivalent to performing the following series of three rotations about the axes of the reference basis:
\begin{equation}\label{eq:euler}
\tensor R=\tensor R^z(\phi)\cdot\tensor R^y(\theta)\cdot\tensor R^z(\psi)
\end{equation}
The matrices associated with the axial rotation about axes $y$ and $z$ are of course, respectively:
\begin{equation}
\tensor R^y(\theta)=
\left(\begin{matrix}
\cos\theta& 0&\sin\theta\\
0&1&0\\
-\sin\theta&0& \cos\theta
\end{matrix}\right)
\end{equation}
and
\begin{equation}\label{eq:Rz}
\tensor R^z(\phi)=
\left(\begin{matrix}
\cos\phi& -\sin\phi& 0\\
\sin\phi& \cos\phi& 0\\
0&0&1
\end{matrix}\right)
\end{equation}

Explicit expressions for the matrices $\tttt{D}^y(\theta)$ and $\tttt{D}^z(\phi)$ were derived in~\cite{Lemaitre2015}, and read:
\begin{equation}\label{eq:Dy}
\begin{split}
\tttt{D}^y(\theta)\!=\!\!
\left(\begin{matrix}
\hspace{1mm}1\hspace{1mm}& 0& 0& 0& 0& \hspace{3mm}0\hspace{3mm}\\
\hspace{1mm}0\hspace{1mm}&\!\!\frac{3}{2}\!\cos^2\theta\!-\!\frac{1}{2}\!\!\!& 0&\!\!\!\! -\frac{\sqrt{3}}{2}\!\sin2\theta\!\!& 0& \!\!\frac{\sqrt{3}}{2}\!\sin^2\theta\!\!\\
\hspace{1mm}0\hspace{1mm}& 0& \cos\theta& 0&\!\!\!\!\! -\sin\theta&  0\\
\hspace{1mm}0\hspace{1mm}&\frac{\sqrt{3}}{2}\!\sin2\theta&  0& \cos2\theta& 0&\!\! -\frac{1}{2}\!\sin2\theta\\
\hspace{1mm}0\hspace{1mm}& 0& \sin\theta& 0& \cos\theta & 0\\
\hspace{1mm}0\hspace{1mm}&\frac{\sqrt{3}}{2}\!\sin^2\theta&  0& \frac{1}{2}\!\sin2\theta& 0&\!\!\frac{1}{2}\!\cos^2\theta\!+\!\frac{1}{2}\,\\
\end{matrix}\right)\\
\end{split}
\end{equation}
and
\begin{equation}\label{eq:Dz}
\tttt{D}^z(\phi)\!=\!\!
\left(\begin{matrix}
\hspace{3mm}1\hspace{3mm}&\hspace{3mm} 0& 0& 0& 0& 0\\[2pt]
\hspace{3mm}0\hspace{3mm}&\hspace{3mm} 1& 0& 0& 0& 0\\[2pt]
\hspace{3mm}0\hspace{3mm}&\hspace{3mm} 0& \cos\phi& \sin\phi&  0&0\\[2pt]
\hspace{3mm}0\hspace{3mm}&\hspace{3mm} 0& -\sin\phi& \cos\phi & 0& 0\\[2pt]
\hspace{3mm}0\hspace{3mm}&\hspace{3mm} 0& 0& 0& \cos2\phi& \sin2\phi\\[2pt]
\hspace{3mm}0\hspace{3mm}&\hspace{3mm} 0& 0& 0&\!\!\!\! -\sin2\phi& \cos2\phi\\[2pt]
\end{matrix}\right)\\[2pt]
\end{equation}
These expressions permit to write the $\tttt{D}$ matrix of a general rotation $\tensor R=\tensor R^z(\phi)\cdot\tensor R^y(\theta)\cdot\tensor R^z(\psi)$ as:
\begin{equation}\label{eq:d:general}
\tttt{D}=\tttt{D}^z(\phi)\cdot\tttt{D}^y(\theta)\cdot\tttt{D}^z(\psi)
\end{equation}

Let us recall a few notions from group theory. The matrices $\tensor R^z$ and $\tttt{D}^z$ account for the action of axial rotations about the $z$ axis on respectivelly vectors and symmetric rank-2 tensors. They are called \emph{representations} of axial rotations about $z$. The block-diagonal structure of matrix $\tensor R^z(\phi)$ in Eq.~(\ref{eq:Rz}) shows that, as is well known, the $z$ axis and the $(x,y)$ plane are invariant under these axial rotations. The restriction of $\tensor R^z(\phi)$ on the $(x,y)$ plane, a $2\times2$ matrix, is called a subrepresentation. It appears to be a 2D rotation. Since 2D rotation do not present any invariant subspace, the subrepresentation of $\tensor R^z(\phi)$ on $(x,y)$ is said irreducible. The restriction of $\tensor R^z(\phi)$ to the $z$ axis is also an irreducible subrepresentation, albeit a trivial one, since it is just the identity. The block structure of the matrix $\tensor R^z$ hence corresponds to its decomposition into irreducible subrepresentations.

The decomposition of $\tttt{D}^z$ into irreducible subrepresentations is immediately visible from its block structure: it leaves invariant the axes $\sigma_1$ and $\sigma_2$, as well as the planes $(\sigma_3,\sigma_4)$ and $(\sigma_5,\sigma_6)$, and its restrictions on the two latter planes are irreducible (they are 2D rotations). The operator $\tttt{D}^y(\theta)$ should also present two invariant lines and two invariant planes. But this is not immediately visible from its matrix structure, which only shows that the subspaces corresponding to the coordinates $\sigma_1$, $(\sigma_2,\sigma_4,\sigma_6)$, and $(\sigma_3,\sigma_5)$ are invariant. Obviously, the subspace $(\sigma_2,\sigma_4,\sigma_6)$ must split further into one invariant line and an invariant plane, but we do not need to identify them.

Concerning $\tttt{D}$, it is obvious that it leaves $\sigma_1$ invariant (pressure is rotation-independent). It is less evident, but will appear crucial that its restriction on $(\sigma_2,\ldots,\sigma_6)$ is irreducible. This holds because any invariant subspace of $\tttt{D}$ must be invariant by both $\tttt{D}^y$ and $\tttt{D}^z$, and none of the invariant subspaces $\sigma_2$, $(\sigma_3,\sigma_4)$ and $(\sigma_5,\sigma_6)$, of $\tttt{D}^z$ is invariant by $\tttt{D}^y$.

\subsubsection{CT and RT autocorrelations}

The stress autocorrelation function of Eq.~(\ref{def:C}) is defined between stress components that are all computed with respect to the reference frame $\basis$. In an isotropic medium, we anticipate that it should be more natural to write the correlation function between stress components that are defined with respect to the basis of spherical coordinates for any difference vector $\vec r$.

Given an arbitrary material point $\vec r$, of spherical coordinates $(r,\theta,\phi)$, we thus define the direction vector $\uvec r = \vec r/r$ and consider the basis $\basis^{\uvec r}=(\vec e_\theta,\ \vec e_\phi,\ \vec e_r=\uvec r)$ with, as usual:
\begin{equation}
\vec e_\theta=\left|
\begin{matrix}
&\!\!\cos\theta\,\cos\phi\\
&\!\!\cos\theta\,\sin\phi\\
&\!\!\!\!\!\!-\sin\theta
\end{matrix}
\right.
\quad
\vec e_\phi=\left|
\begin{matrix}
&\!\!-\sin\phi\\
&\!\!\cos\phi\\
&0
\end{matrix}
\right.
\quad
\vec e_r=\left|
\begin{matrix}
&\!\!\sin\theta\,\cos\phi\\
&\!\!\sin\theta\,\sin\phi\\
&\!\!\cos\theta
\end{matrix}
\right.
\end{equation}
The tesseral tensor components of stress in $\basis^{\uvec r}$ are denoted $\utilde\sigma^{\uvec r}$. They are, by definition, the same as in Eq.~(\ref{eq:stress}), up to the $(x,y,z)\to(\theta,\phi,r)$ transformation and read explicitly:
\begin{equation}
\label{eq:stress:rst}
\begin{split}
\sigma_1^{\uvec r}&=-\frac{1}{\sqrt{3}}\,\left(\sigma_{\theta\theta}+\sigma_{\phi\phi}+\sigma_{rr}\right)\\
\sigma_2^{\uvec r}&=-\frac{1}{\sqrt{6}}\,\left(\sigma_{\theta\theta}+\sigma_{\phi\phi}-2\,\sigma_{rr}\right)\\
\sigma_3^{\uvec r}&=\sqrt{2}\,\sigma_{\phi r}\\
\sigma_4^{\uvec r}&=\sqrt{2}\,\sigma_{\theta r}\\
\sigma_5^{\uvec r}&=\sqrt{2}\,\sigma_{\theta\phi}\\
\sigma_6^{\uvec r}&=\frac{1}{\sqrt{2}}\,\left(\sigma_{\theta\theta}-\sigma_{\phi\phi}\right)\\
\end{split}
\end{equation}

Let us recall that, in our convention, the rotation $\tensor R^{\uvec r}$ associated with the $\basis\to\basis^{\uvec r}$ change of basis is such that $(\vec e_\theta,\ \vec e_\phi,\ \vec e_r)=(\tensor R^{\uvec r})^T\cdot(\vec e_x,\vec e_y,\vec e_z)$. Note also the chosen order for basis vectors: $\basis^{\uvec r}$ is produced by operating on $(\vec e_x,\vec e_y,\vec e_z)$ a rotation by $\phi$ about the $z$ axis followed by a rotation by $\theta$ about the new $y$ axis, i.e. a ZYZ Euler rotation of angles $(\phi, \theta, 0)$. Therefore, $(\tensor R^{\uvec r})^T=\tensor R^z(\phi)\cdot\tensor R^y(\theta)$, so that:
\begin{equation}\label{eq:Rr}
\tensor R^{\uvec r}=(\tensor R^z(\phi)\cdot\tensor R^y(\theta))^T
\end{equation}
The matrix $\tttt{D}^{\uvec r}$, which performs the $\basis\to\basis^{\uvec r}$ change of basis on tesseral components, i.e. such that,
\begin{equation}\label{eq:cstors:sigma}
\utilde\sigma^{\uvec r}=\tttt{D}^{\uvec r}\cdot\utilde\sigma
\end{equation}
is hence:
\begin{equation}
\tttt{D}^{\uvec r}=(\tttt{D}^z(\phi)\cdot\tttt{D}^y(\theta))^T
\end{equation}
Its fully explicit form can be found in~\cite{Lemaitre2015}.\\

In the following, the stress correlation matrix defined in Eq.~(\ref{def:C}) in the reference basis will be refered to as the \emph{Cartesian tesseral} (or CT) stress correlation. The \emph{Radial tesseral} (or RT) stress autocorrelation is defined, for any non-zero $\vec r$, as the field:
\begin{equation}\label{eq:C:radial}
\mathring{\uttilde{C}}(\vec r)=\langle\utilde \sigma^{\uvec r}(\vec r_0+\vec r)\,\utilde \sigma^{\uvec r}(\vec r_0)\rangle_c
\end{equation}
For any pair $\vec r_0$ and $\vec r_0+\vec r$, it is the correlation matrix between the stress components in basis $\basis^{\uvec r}$. Like $\uttilde{C}$ it is invariant under translations and spatial inversion symmetry, hence it is independent on $\vec r_0$. It is also a symmetric matrix. Using equation~(\ref{eq:cstors:sigma}), we observe that the RT and CT forms of the stress autocorrelation are related by the following identity,
\begin{equation}
\label{eq:cstors:C}
\mathring{\uttilde{C}}(\vec r) =\tttt{D}^{\uvec r}\cdot\uttilde{C}(\vec r)\cdot(\tttt{D}^{\uvec r})^T
\end{equation}
at any non-zero $\vec r$.

\section{Conditions and consequences of material isotropy}
\label{sec:materialisotropy}

The main elements of our formalism are now laid down. We now turn to the core question of this work, of understanding how material isotropy (in this Section) and mechanical balance (in the next one) constrain the spatial structure of the stress autocorrelation function. Our theoretical investigation, of course, concerns the case of an infinite medium.

In this section, we write the explicit form of the stress autocorrelation that guarantees its agreement with material isotropy, i.e. with equation~(\ref{eq:iso:condition}). While it essentially follows the ideas of~\cite{Lemaitre2015} our presentation is simpler and lays the groundwork for the follow up calculations.

\subsubsection{Stress correlation and material isotropy}

Let us view a stress autocorrelation field as a function $\uttilde{C}(\vec r)$ in the lab frame $\basis$. In $\basis^{\tensor R}$, the same field is a different function, ${\uttilde{C}}'$ of the running coordinate, namely:
\begin{equation}
{\uttilde{C}}'(\vec r') =\tttt{D}\cdot\uttilde{C}\left(\vec r\right)\cdot\tttt{D}^T
\mathcal{}\end{equation}
with $\vec r=\tensor R^T\cdot\vec r'$.

Material isotropy is, by definition, the property that the functional form ${\uttilde{C}}'(\vec r')$ is identical in any basis. This amounts to the conditions that ${\uttilde{C}}'(\vec r')=\uttilde{C}(\vec r')$ for any $\basis^{\tensor R}$ and $\vec r'$. Material isotropy is therefore equivalent to requiring that:
\begin{equation}
\label{eq:iso:condition}
{\uttilde{C}}(\tensor R\cdot\vec r) =\tttt{D}\cdot\uttilde{C}\left(\vec r\right)\cdot\tttt{D}^T
\end{equation}
for any $\vec r$ and ${\tensor R}$.

The above relation implies that the spatial structure of the stress autocorrelation field is tightly constrained. Take indeed any non-zero vector $\vec r$: when $\tensor R$ spans the set of all rotations, $\tensor R\cdot\vec r$ covers (multiple times) the sphere of radius $r=\|\vec r\|$. The above relation implies that the value of ${\uttilde{C}}$ at the single point $\vec r$ fixes it everywhere on the radius-$r$ sphere. By extension, the value of $\uttilde{C}$ along a single, arbitrary axis fixes it everywhere in space.

In the following, we would like to identify an explicit expression for $\uttilde{C}$ that guarantees its consistency with material isotropy. Such expressions are well-known for scalar or vector fields. The problem we face here is that we cannot just take any value of $\uttilde{C}\left(\vec r\right)$ along an arbitrary axis, say for all $\vec r=r\vec e_z$, an use Eq.~(\ref{eq:iso:condition}) as an explicit definition of the function everywhere in space, because we are not guaranteed that any two rotations $\tensor R_1$ and $\tensor R_2$ that give the same $\vec r'=\tensor R_1\cdot (r\,\vec e_z)=\tensor R_2\cdot (r\,\vec e_z)$ will also give the same left hand side in Eq.~(\ref{eq:iso:condition}). Hence, we are not guaranteed to construct a mono-valued function in this way. Equation~(\ref{eq:iso:condition}) contains certain implicit conditions on $\uttilde{C}$ that must be spelled out.


\subsubsection{Explicit form of the autocorrelation matrix in real space}

Here, we identify two consequences of material isotropy, that will later prove to be sufficient.\\

First, observe that in an isotropic medium:
\begin{equation}\label{eq:independent}
\mathring{\uttilde{C}}(\vec r)=\mathring{\uttilde{C}}(r)
\end{equation}
is a radial function. This is obvious because the direction $\uvec r$ used in the definition of $\mathring{\uttilde{C}}$ [Eq.~(\ref{eq:C:radial})] is irrelevant. This property can also be deduced formally from Eq.~(\ref{eq:iso:condition}), by taking the particular case of $\tensor R=\tensor R^{\uvec r}$, which maps $\basis\to\basis^{\uvec r}$ and $\vec r$ onto $\vec r'=\tensor R^{\uvec r}\cdot\vec r=(0,0,r)$, which yields:
\begin{equation}\label{eq:z}
\mathring{\uttilde{C}}(\vec r) = \uttilde{C}(0,0,r)=\uttilde{C}(r\vec e_z)
\end{equation}

An important consequence of~(\ref{eq:independent}) emerges as soon as we invert Eq.~(\ref{eq:cstors:C}) and write:
\begin{equation}
\label{eq:cstors:C:inverse}
\uttilde{C}(\vec r)=\left(\tttt{D}^{\uvec r}\right)^T\cdot\mathring{\uttilde{C}}(r)\cdot\tttt{D}^{\uvec r}
\end{equation}
Since here (in an isotropic medium) the RT correlation is a function of $r$ only, this equation demonstrates that all angular-dependencies of $\uttilde{C}(\vec r)$ come solely from the $\tttt{D}^{\uvec r}$ matrix, i.e. from trivial contributions accounting for tensor rotations~\cite{Lemaitre2015,LevashovStepanov2016}.\\

A second consequence of material isotropy emerges as soon as one take $\tensor R=\tensor R^z(\psi)$ and $\vec r=r\vec e_z$ in Eq.~(\ref{eq:iso:condition}), which yields [using Eq.~(\ref{eq:z})]:
\begin{equation}\label{eq:axial}
\mathring{\uttilde{C}}(r)=\tttt{D}^z(\psi)\cdot\mathring{\uttilde{C}}(r)\cdot\tttt{D}^z(-\psi)
\end{equation}
This equation expresses that $\mathring{\uttilde{C}}(r)$ is invariant under any axial rotation around $\uvec r$. Let us recall that $\tttt{D}^z(\theta)$ is block diagonal. As seen on Eq.~(\ref{eq:Dz}), it presents three $2\times2$ blocks, the first being the identity matrix and the next two corresponding to 2D rotations by $\phi$ and $2\phi$, respectively. Schur's first lemma states that to be invariant under all such rotations, $\mathring{\uttilde{C}}(r)$ must present the same block diagonal structure, with the first block arbitrary, and the two others proportional to the $2\times2$ identity matrix; it must hence be of the form
\begin{equation}
  \label{eq:constraint:C}
\mathring{\uttilde{C}}(\vec r)=\left(
  \begin{matrix}
\mathring{C}_1(r)&\mathring{C}_2(r)&      0&      0&      0&      0\\
\mathring{C}_2(r)&\mathring{C}_3(r)&      0&      0&      0&      0\\
      0&      0&\mathring{C}_4(r)&      0&      0&      0\\
      0&      0&      0&\mathring{C}_4(r)&      0&      0\\
      0&      0&      0&      0&\mathring{C}_5(r)&      0\\
      0&      0&      0&      0&      0&\mathring{C}_5(r)
  \end{matrix}
\right)
\end{equation}
with components $\mathring{C}_i$ that are functions of $r$ only. Remarkably, the autocorrelation matrix is fixed by only five scalar functions of $r$.\\

We have shown that Eqs.~(\ref{eq:independent}) and~(\ref{eq:constraint:C}) are two consequences of material isotropy. We now show that they are sufficient, together, to imply material isotropy, i.e. Eq.~(\ref{eq:iso:condition}). To do so, we assume they both hold, and consider an arbitrary point $\vec r$ and an arbitrary rotation $\tensor R$. We introduce the rotation $\tensor R^{\tensor R\cdot\uvec r}$ associated with the $\basis\to\basis^{\tensor R\cdot\uvec r}$ frame change. By definition, $\tensor R^{\tensor R\cdot\uvec r}\cdot\tensor R\cdot\uvec r=\vec e_z$ and also $\tensor R^{\uvec r}\cdot\uvec r=\vec e_z$. Therefore, $\tensor R^{\tensor R\cdot\uvec r}\cdot\tensor R\cdot(\tensor R^{\uvec r})^T$ leaves $\vec e_z$ invariant: it is an axial rotation about $\vec e_z$ by some angle $\psi$, and we may write $\tensor R^{\tensor R\cdot\uvec r}=\tensor R^z(\psi)\cdot\tensor R^{\uvec r}\cdot\tensor R^T$. It now suffices to compute the lhs of Eq.~(\ref{eq:iso:condition}) using~(\ref{eq:cstors:C:inverse}): $\uttilde{C}(\tensor R\cdot\vec r)=\left(\tttt{D}^{\tensor R\cdot\uvec r}\right)^T\cdot\mathring{\uttilde{C}}(r)\cdot\tttt{D}^{\tensor R\cdot\uvec r}=\tttt{D}\cdot(\tttt{D}^{\uvec r})^T\cdot\tttt{D}^{z}(-\psi)\cdot\mathring{\uttilde{C}}(r)\cdot\tttt{D}^{z}(\psi)\cdot\tttt{D}^{\uvec r}\cdot\tttt{D}^T$; using Eq.~(\ref{eq:axial}) and then Eq.~(\ref{eq:cstors:C:inverse}), we recover Eq.~(\ref{eq:iso:condition}), which concludes the proof.

The above argument establishes that a stress correlation function is consistent with material isotropy if and only if its RT expression $\mathring{\uttilde{C}}$ is:
\begin{compactenum}
\item spatially isotropic (radial) [Eq.~(\ref{eq:independent})]
\item of the matrix form described in Eq.~(\ref{eq:constraint:C}).
\end{compactenum}

\subsubsection{Stress correlations in Fourier space}
\label{sec:fourier}
The Fourier transform $\mathcal{F}[f]\equiv\widehat{f}$ of a function $f$ of the infinite continuum is defined as usual as:
\begin{equation}\label{eq:fourier}
\hat f({\vec k})\equiv\int\d^3\vec r\,f(r)\,e^{-i\vec k\cdot\vec r}
\end{equation}
with the inverse formula:
\begin{equation}\label{eq:fourier:cont}
f(\vec r)=\frac{1}{(2\pi)^3}\,\int\d^3{\vec k}\,e^{i\vec k\cdot\vec r}\ \hat f({\vec k})
\end{equation}

The Fourier transform of $\uttilde{C}(\vec r)$ is the function:
\begin{equation}
\label{eq:Ctt:fourier:CT}
\uttilde{\widehat{C}}({\vec k})=\frac{1}{(2\pi)^3}\,\left\langle\utilde{\widehat{\sigma}}({\vec k})\,\left(\utilde{\widehat{\sigma}}({\vec k})\right)^*\right\rangle_c
\end{equation}
with $^*$ the complex conjugate.
Its RT form is the matrix:
\begin{equation}
\label{eq:Ctt:fourier:RT}
\uttilde{\mathring{\widehat{C}}}({\vec k}) =\frac{1}{(2\pi)^3}\,\left\langle\utilde{\widehat{\sigma}}^{\uvec k}({\vec k})\,\left(\utilde{\widehat{\sigma}}^{\uvec k}({\vec k})\right)^*\right\rangle_c
\end{equation}
where $\utilde{\widehat{\sigma}}^{\uvec k}({\vec k})\equiv\mathcal{D}^{\uvec k}\cdot\utilde{\widehat{\sigma}}({\vec k})$. Since $\mathcal{D}^{\uvec k}$ is real, the relation between the CT and RT forms reads exactly the same as in real space [Eq.~(\ref{eq:cstors:C})]:
\begin{equation}
\label{eq:Ctt:fourier}
\uttilde{\mathring{\widehat{C}}}({\vec k}) =\tttt{D}^{\uvec k}\cdot\uttilde{\widehat{C}}({\vec k})\cdot(\tttt{D}^{\uvec k})^T
\end{equation}
Both fields $\uttilde{\widehat{C}}$ and $\uttilde{\mathring{\widehat{C}}}$ are real-valued thanks to spatial inversion symmetry in real space; it then immediately appears from their definitions, Eqs.~(\ref{eq:Ctt:fourier:CT}) and~(\ref{eq:Ctt:fourier:RT}), that they are both symmetric matrices; they are also invariant under the $\vec k\to-\vec k$ inversion, because the real space stress correlation is real-valued. It should finally be noted that, the diagonal elements of both $\uttilde{{\widehat{C}}}({\vec k})$ and $\uttilde{\mathring{\widehat{C}}}({\vec k})$ are non-negative: this is an instance of the Wiener-Khintchine theorem, and comes out immediately by inspection of their definitions~(\ref{eq:Ctt:fourier:CT}) and~(\ref{eq:Ctt:fourier:RT}).

For the same reasons of symmetry as in real space, $\uttilde{\mathring{\widehat{C}}}({\vec k})$ is $\uvec k$-independent in an isotropic medium, and since axial symmetry applies just the same, it presents, for any non-zero $\vec k$, the matrix structure of Eq.~(\ref{eq:constraint:C}):
\begin{equation}
  \label{eq:constraint:k}
\begin{split}
&\forall\vec k\ne\vec 0\\
&\mathring{\uttilde{\widehat{C}}}({\vec k})=\left(
  \begin{matrix}
\mathring{\widehat{C}}_1(k)&\mathring{\widehat{C}}_2(k)&                  0&                0&                0&                0\\
\mathring{\widehat{C}}_2(k)&\mathring{\widehat{C}}_3(k)&                  0&                0&                0&                0\\
                0&                0&\mathring{\widehat{C}}_4(k) &                0&                0&                0\\
                0&                0&                0&\mathring{\widehat{C}}_4(k) &                0&                0\\
                0&                0&                0&                0&\mathring{\widehat{C}}_5(k) &                0\\
                0&                0&                0&                0&                0&\mathring{\widehat{C}}_5(k)
\end{matrix}
\right)
\end{split}
\end{equation}
which involves five scalar functions $\mathring{\widehat{C}}_i(k)$ of the wavevector amplitude.

Let us pay attention to the fact that 
except for $\mathring{\widehat{C}}_1$ the pressure autocorrelation (see details below), the functions $\mathring{\widehat{C}}_i(k)$ are \emph{not} the Fourier transforms of the functions $\mathring{C}_i(r)$ of Eq.~(\ref{eq:constraint:C}) [$\mathring{\widehat{C}}_i\ne\widehat{\mathring{C}}_i$ in general]. It is $\uttilde{\widehat{C}}$ which is the Fourier transform of $\uttilde{C}$. It may help to picture the relation between CT, RT fields in real and Fourier space using the following diagram:
\begin{equation}\label{diag}
\begin{CD}
  \uttilde{C}@>\qquad\mathcal{F}\qquad>>&\uttilde{\widehat{C}}\\
  @VV \text{Eq.~(\ref{eq:cstors:C})}V&@VV \text{Eq.~(\ref{eq:Ctt:fourier})}V\\
  \mathring{\uttilde{{C}}}@.&\mathring{\uttilde{\widehat{C}}}
\end{CD}
\end{equation}
The main difficulty of our analysis is that isotropy (and as we will shortly see mechanical balance) are better expressed using the RT forms $\mathring{\uttilde{{C}}}$ and $\mathring{\uttilde{\widehat{C}}}$. However, as the above diagram emphasizes, there is no direct relation between these two functions, i.e. between the $\mathring{C}_i(r)$ and $\mathring{\widehat{C}}_i(k)$'s. 

\subsubsection{Isotropic and anisotropic parts of CT correlations}
\label{sec:isotropy}

It is now clear that, in an isotropic medium, the stress autocorrelation function $\uttilde{C}$ presents spatial anisotropies confered by the matrix products appearing in Eq.~(\ref{eq:cstors:C:inverse}), which capsulize the trivial tensor rotations between the radial and Cartesian frames. It does not mean that every term contributing to $\uttilde{C}$ is anisotropic, however. The clearest counterexample is the pressure autocorrelation, which is the component $C_{11}$.
It should be emphasized that it always verifies (independently of material isotropy):
\begin{equation}\label{eq:pressure}
\mathring{C}_{11}=C_{11}
\end{equation}
because pressure is a rotation-invariant (scalar) quantity. It is this equation which guarantees that the pressure autocorrelation ${C}_{11}$ is radial in an isotropic medium---since $\mathring{C}_{11}$ then is.

Consider also the trace of $\uttilde{C}$, denoted $\trace[\uttilde{C}]$. We may write, for any $\vec r\ne\vec0$:
\begin{equation}\label{eq:trace}
\trace[\mathring{\uttilde C}]=\trace[\tttt{D}^{\uvec r}\cdot \uttilde C\cdot (\tttt{D}^{\uvec r})^T]=\trace[\uttilde C]
\end{equation}
which is similar to Eq.~(\ref{eq:pressure}). By the same line of argument as for the pressure autocorrelation, this equation guarantees that $\trace[\uttilde{C}]$ is radial in an isotropic medium since $\trace[\mathring{\uttilde{C}}]$ is. We thus have identified two contributions to $\uttilde{C}$ that are radial in an isotropic medium. Showing that they are the only ones necessitates a slightly formal argument with the introduction of the notion of isotropic tensor.\\

An \emph{isotropic tensor} is defined as a tensor which is invariant under any rotation. For a minor-symmetric fourth order tensor ${\uttilde{C}}$ (a $6\times6$ matrix in our formalism), tensor isotropy thus amounts to requiring:
\begin{equation}
  {\uttilde{C}}=\tttt{D}^T\cdot{\uttilde{C}}\cdot\tttt{D}
\end{equation}
for any $\tttt{D}$ representing a rotation. As explained after Eq.~(\ref{eq:d:general}), arbitrary $\tttt{D}$ matrices have a block diagonal structure with a $1\times1$ block, equal to 1, because pressure is invariant, and a $5\times5$ block that operates a rotation on the subspace $(\sigma_2,\ldots,\sigma_6)$, which correspond to their two irreducible subrepresentations. Now, Schur's first lemma (again) states that to be invariant under all such rotations, ${\uttilde{C}}$ has to present the same block-diagonal structure and must be proportional to the identity on each block corresponding to a (non-trivial) invariant subspace. It follows that ${\uttilde{C}}$ is isotropic iff it is of the form:
\begin{equation}
  \label{eq:tensor:isotropic}
\begin{split}
{\uttilde{C}}&=\left(
  \begin{matrix}
C_0&      0&      0&      0&      0&      0\\
   0&  C_0'&      0&      0&      0&      0\\
   0&      0&  C_0'&      0&      0&      0\\
   0&      0&      0&  C_0'&      0&      0\\
   0&      0&      0&      0&  C_0'&      0\\
   0&      0&      0&      0&      0&   C_0'
  \end{matrix}
\right)
\end{split}
\end{equation}
with two coefficients $C_0$ and $C_0'$.\\

Let us now consider an arbitrary $6\times6$ matrix $\uttilde{C}$. We define its \emph{isotropic part} as the matrix $\Iso\left[{\uttilde{C}}\right]$ of the form~(\ref{eq:tensor:isotropic}) with the coefficients:
\begin{equation}\label{eq:trace:cs}
\begin{split}
C_0&\equiv C_{11}\\
C_0'&\equiv\frac{1}{5}\,\sum_{a=2}^6 C_{aa}=\frac{1}{5}\,\left(\trace\left[\uttilde{C}\right]-C_{11}\right)
\end{split}
\end{equation}
Since $\Iso\left[\Iso\uttilde{C}\right]=\Iso\uttilde{C}$, the operation $\Iso$ is a projection onto the set of isotropic tensors. The coefficients $C_0$ and $C_0'$ (or their combinations) are therefore the only rotation-invariant contributions to $\uttilde{C}$.

Observe that the word \emph{isotropy} is used in reference to different notions that must be carefully distinguished: material, spatial, and tensorial. It should be noticed, in particular, that the qualifier ``isotropic'' in ``isotropic part'' refers to tensor isotropy and does not imply spatial isotropy. In fact, for a general stress autocorrelation field $\uttilde{C}(\vec r)$ (general meaning, without assuming material isotropy), the isotropic tensor part $\Iso[\uttilde{C}](\vec r)$ is not necessarily spatially isotropic (neither is the pressure autocorrelation).\\

The operation $\Iso$ presents two properties that will be of great help. First, it obviously commutes with the Fourier transform:
\begin{equation}\label{eq:isoft}
\Iso\left[\mathcal{F}[\uttilde{C}]\right]=\mathcal{F}\left[\Iso[\uttilde{C}]\right]
\end{equation}
which guarantees that there is no ambiguity when we write $\widehat{C}_0$ and $\widehat{C}_0'$ the two coefficients of $\Iso\uttilde{\widehat{C}}$ such as in Eqs~(\ref{eq:tensor:isotropic}) and~(\ref{eq:trace:cs}): they are the Fourier transforms of the corresponding functions ${C}_0$ and ${C}_0'$ in $\Iso[\uttilde{C}](\vec r)$.
Second, at any point $\vec r\ne\vec0$, we may write:
\begin{equation}\label{eq:iso:csrs}
\Iso\left[{\uttilde{C}}(\vec r)\right]=\Iso\left[\mathring{\uttilde{C}}(\vec r)\right]
\end{equation}
(the origin is excluded simply because the RT form is not defined there). This is indeed an obvious consequence of Eqs.~(\ref{eq:pressure}) and~(\ref{eq:trace}), and thus holds in all generality.
The same, of course, applies in Fourier space:
\begin{equation}\label{eq:iso:csrs:fourier}
\Iso\left[{\uttilde{\widehat{C}}}(\vec k)\right]=\Iso\left[\mathring{\uttilde{\widehat{C}}}(\vec k)\right]
\end{equation}
for any $\vec k\ne\vec 0$.


Our interest lies not in arbitrary stress correlation functions but is those that are consistent with material isotropy. In that case, as we have previously shown, the RT form $\mathring{\uttilde{C}}(\vec r)$ is radial. Its isotropic part $\Iso\mathring{\uttilde{C}}(\vec r)$ is hence radial too, and so does $\Iso{\uttilde{C}}(\vec r)$ in view of Eq.~(\ref{eq:iso:csrs}). We therefore conclude that: \emph{the isotropic (tensor) part of a materially isotropic tensor field is spatially isotropic.}
Additionally, using Eq.~(\ref{eq:constraint:C}), we may then write:
\begin{equation}\label{eq:trace:rs}
\begin{split}
C_0(r) &= \mathring{C}_{1}(r)\\
C_0'(r)&=\frac{\mathring{C}_3(r)+2\mathring{C}_4(r)+2\mathring{C}_5(r)}{5}
\end{split}
\end{equation}
which will prove to be very useful. The same relation hold of course in Fourier space:
\begin{equation}\label{eq:trace:rs:Fourier}
\begin{split}
\widehat{C}_0(k) &=\mathring{\widehat{C}}_{1}(k)\\
\widehat{C}_0'(k)&=\frac{\mathring{\widehat{C}}_3(k)+2\mathring{\widehat{C}}_4(k)+2\mathring{\widehat{C}}_5(k)}{5}
\end{split}
\end{equation}
Since $\Iso\left[\mathcal{F}[\uttilde{C}]\right]=\mathcal{F}\left[\Iso[\uttilde{C}]\right]$ these two pairs of equations fix rather simple relations between the $\mathring{C}_i$ and $\mathring{\widehat{C}}_i$ functions.

Let us remark that our introduction of the RT and CT forms, our identification of 6 components for the stress correlations in an isotropic medium, and finally our definition of the isotropic part, have led us to attribute different symbols that all correspond to the pressure autocorrelation:
\begin{equation}
C_{11}=\mathring{C}_{11}=\mathring{C}_{1}=C_0
\end{equation}
and likewise in Fourier space.

To conclude this discussion, let us show that $\Iso[\uttilde{C}]$ captures all the radial contributions to the entire correlation function $\uttilde{C}$ in a materially isotropic medium. To do so, let us define as $\uttilde{C}^{\rm iso}$ all the spatially radial contributions to $\uttilde{C}$. These radial terms must be identical in all frames, hence, must be consistent with material isotropy. Radial symmetry guarantees that, for any rotation $\tensor R$, $\uttilde{C}^{\rm iso}(\tensor R\cdot\vec r)=\uttilde{C}^{\rm iso}(\vec r)$; and equation~(\ref{eq:iso:condition}), then implies that $\uttilde{C}^{\rm iso}$ is invariant under all rotations, i.e. that it is an isotropic tensor. It can hence only identify with $\Iso[\uttilde{C}]$. It follows that, if we denote $\uttilde{C}^{\rm iso}=\Iso[\uttilde{C}]$ and write $\uttilde{C}=\uttilde{C}^{\rm iso}+\uttilde{C}^{\rm ani}$, we split $\uttilde{C}$ into isotropic and anisotropic parts in terms of both tensorial and spatial isotropy.

\subsubsection{Fluctuations of the sphere-averaged stress}
\label{sec:thermo}

We will show here that, in an isotropic medium, the two functions $C_0$ and $C_0'$ are intricately related to the decay of the fluctuations of the window-averaged stress. To proceed, let us consider the average stress on a spherical observation window of radius $R$, centered around an arbitrary point $\vec r$:
\begin{equation}\label{def:stress:cg}
\overline{\utilde\sigma}(\vec r;R)=\frac{1}{\Omega_R}\int_{\|\vec r-\vec r'\|<R}\!\!\d^3\vec r'\ \utilde{\sigma}(\vec r')
\end{equation}
We will refer to this observable as the sphere-averaged stress. Its correlations and fluctuations are captured by the matrix:
\begin{equation}
\begin{split}
\uttilde{J}(R)&\equiv\langle\,\overline{\utilde\sigma}(\vec r;R)\,\overline{\utilde\sigma}(\vec r;R)\,\rangle\\
&=\frac{1}{\Omega_R^2}\ \int_{r_1<R}\!\!\d^3\vec r_1\int_{r_2<R}\!\!\d^3\vec r_2 \ \uttilde C\left(\vec r_2-\vec r_1\right)
\end{split}
\end{equation}
with $\Omega_R=4\pi\,R^3/3$ the volume of the sphere.

It is easy to see that $\uttilde{J}$ is an isotropic tensor: pick an arbitrary rotation $\tensor R$ and introduce the variables $\vec r_1'=\tensor R\cdot\vec r_1$ and $\vec r_2'=\tensor R\cdot\vec r_2$; since rotations have a determinant equal to unity, using~(\ref{eq:iso:condition}) it comes:
\begin{equation}
\uttilde J(R)=\tttt{D}^T\cdot\uttilde J(R)\cdot\tttt{D}
\end{equation}
Since $\uttilde{J}$ is an isotropic tensor, it is of the form~(\ref{eq:tensor:isotropic}), which implies that:
\begin{compactitem}
\item[(i)] all cross correlations between the components of the sphere-averaged stress vanish
\item[(ii)] all the fluctuations of all sphere-averaged deviatoric components are identical.
\end{compactitem}
Using the property $\uttilde J=\Iso\uttilde J$, we may now write
\begin{equation}\label{eq:JR}
\begin{split}
\uttilde J(R)&=\Iso\uttilde J(R)\\
&=\frac{1}{\Omega_R^2}\ \int_{r_1<R}\!\!\d^3\vec r_1\int_{r_2<R}\!\!\d^3\vec r_2 \ \uttilde C^{\rm iso}\left(\|\vec r_2-\vec r_1\|\right)
\end{split}
\end{equation}
which shows that the pressure and deviatoric stress fluctuations depend, respectivelly, on the radial functions $C_0$ and $C_0'$ that constitute $\uttilde C^{\rm iso}$.\\

To understand in more detail the relation between sphere-averaged stress fluctuations and the stress autocorrelation function, it is convenient, following Refs.~\cite{TorquatoStillinger2003,Torquato2016,Torquato2018}, to introduce the window indicator function,
\begin{equation}
w(\vec r; R) = \left\{
\begin{aligned}
  &1 \qquad \textrm{if}\ \|\vec r\|<R\\
  &0\qquad \textrm{otherwise}
\end{aligned}\right.
\end{equation}
and the \emph{scaled intersection volume function}~\cite{Torquato2018}:
\begin{equation}\label{def:alpha}
  \alpha(\vec r; R) = \frac{1}{\Omega_R}\int\d^3\vec r_0\ w(\vec r_0;R)\,w(\vec r_0+\vec r;R)
\end{equation}
using which Eq.~(\ref{eq:JR}) may be written~\cite{TorquatoStillinger2003,Torquato2016,Torquato2018}:
\begin{equation}\label{eq:J:alpha:fourier}
\uttilde{J}(R)=\frac{1}{\Omega_R}\ \int\,\frac{\d^3\vec k}{(2\pi)^3}\,\widehat\alpha(\vec k;R)\,\widehat{\uttilde C}^{\rm iso}\left(\vec k\right)
\end{equation}
where the function $\widehat\alpha$ converges to $(2\pi)^3\,\delta(\vec k)$ when $R\to\infty$~\cite{TorquatoStillinger2003,Torquato2018} [see details in Appendix~\ref{sec:integrals}]. Note that, like all diagonal elements of $\uttilde{\widehat{C}}$, the two functions $\widehat{C}_0$ and $\widehat{C}_0'$ are non-negative, which guarantees that the rhs of Eq.~(\ref{eq:J:alpha:fourier}) is non-negative too. The above equation demonstrates that the large-$R$ behavior of $\uttilde{J}(R)$, i.e. of sphere-averaged stress fluctuations, is entirely fixed by the isotropic part of the stress autocorrelation, and especially by its low-$k$ behavior.\\

In all generality, we should envision that the large-$R$ behavior of $\uttilde{J}(R)$ may fit under either of the following three cases:
\begin{compactenum}
\item $\Omega_R\uttilde{J}(R)$ converges to a non-zero constant: this is the normal behavior which, as we will argue below, is expected to apply to liquids or glasses
\item $\Omega_R\,\uttilde{J}(R)$ diverges: a situation that may be encountered near critical points
\item $\Omega_R\,\uttilde{J}(R)$ vanishes: which would correspond to ``hyperuniform stresses''; we are not aware that this behavior is found in any system, but cannot exclude it either.
\end{compactenum}

Window-averaged stress fluctuations are expected to obey the normal scaling in glasses (and liquids). Indeed, when we speak of a glassy state, we consider systems that have a well-defined stress state in the thermodynamic limit, namely a finite pressure and zero deviatoric stresses. This mechanical state is well-defined only if stress is self-averaging, which rules out that $\Omega_R\,\uttilde{J}(R)$ diverges at large $R$. This argument does not exclude the possibility that stress fluctuations may be hyperuniform, but this latter behavior requires a rare degree of structure, and hence cannot be considered as generic in disordered systems. For these reasons, we expect that the normal scaling behavior applies to glasses and liquids and will emphasize this case in our analysis; but we will also discuss other possibilities.

In view of Eq.~(\ref{eq:J:alpha:fourier}), since $\widehat\alpha(\vec k;R)\to(2\pi)^3\,\delta(\vec k)$ when $R\to\infty$, the existence of a finite limit value (zero or non-zero) for $\Omega_R\uttilde{J}(R)$ requires $\uttilde{\widehat{C}}^{\rm iso}(\vec k)$ to be continuous hence to converge in the $\vec k\to\vec 0$ limit. The following property then holds:
\begin{equation}
{\Omega_R}\,\uttilde{J}(R)\xrightarrow[R \to\infty]{}\ \uttilde{\widehat{C}}^{\rm iso}(\vec 0)
\end{equation}
where
\begin{equation}
\begin{split}
  \uttilde{\widehat{C}}^{\rm iso}(\vec 0)&=
  \begin{pmatrix}
\widehat{C}_0(0)&    0&    0&    0&      0&      0\\
   0&\widehat{C}_0'(0)&    0&    0&      0&      0\\
   0&    0&\widehat{C}_0'(0)&    0&      0&      0\\
   0&    0&    0&\widehat{C}_0'(0)&      0&      0\\
   0&    0&    0&    0&\widehat{C}_0'(0)&      0\\
   0&    0&    0&    0&      0&\widehat{C}_0'(0)
  \end{pmatrix}
\end{split}
\end{equation}
This argument applies both to the normal case (1. above) and to the hypothetical hyperuniform stress case (3.).
We have now established that, in these cases, the decay of stress fluctuations is fixed by:
\begin{equation}\label{limit:thermo:2}
\begin{split}
\Omega_R\,\left\langle\,\left(\overline{\sigma}_1(\vec r;R)\right)^2\,\right\rangle&\xrightarrow[R \to\infty]{}\widehat{C}_0(0)\\
\forall a\ne1\quad\Omega_R\,\left\langle\,\left(\overline{\sigma}_a(\vec r;R)\right)^2\,\right\rangle&\xrightarrow[R \to\infty]{}\widehat{C}_0'(0)
\end{split}
\end{equation}

Let us finally note that, since the full correlation matrix $\uttilde{\widehat{C}}(\vec 0)$ is invariant under all rotations at $\vec k=\vec 0$, it is an isotropic tensor at that point so that, as soon as $\uttilde{\widehat{C}}^{\rm iso}(\vec 0)$ is finite, we may also write:
\begin{equation}
\uttilde{\widehat{C}}^{\rm iso}(\vec 0)=\uttilde{\widehat{C}}(\vec 0)
\end{equation}
However, we need to keep in mind that while $\uttilde{\widehat{C}}^{\rm iso}$ is a continuous function at the origin, the full autocorrelation $\uttilde{\widehat{C}}(\vec k)$ is not in general, due to the anisotropic contributions. This will become especially clear in the next section as we will see that certain diagonal components of $\uttilde{\widehat{C}}(\vec k)$ vanish exactly along e.g. the $\vec e_z$ axis, a property called directional hyperuniformity~\cite{Torquato2016}, but not in all directions.

\section{Compounding material isotropy with mechanical balance}
\label{sec:compounding}

We have so far examined the consequences of material isotropy only. The results we have obtained hold for any stress correlation in any isotropic system, irrespective of mechanical balance. They may apply, for example, to stress in liquid configurations. Now, we turn to the complete problem of understanding the structure of stress correlations in systems that are both isotropic and mechanically balanced.

\subsubsection{Complete expression in Fourier space}
\label{sec:constraints}

Let us first examine the consequence of mechanical balance, $i\vec k\cdot\tensor{\widehat\sigma}({\vec k})=\vec 0$, alone. Denoting $(k,\thetak,\phik)$ the polar coordinates in Fourier space, this condition is equivalently written: $\widehat\sigma_{kk}(\vec k)=\widehat\sigma_{k\thetak}(\vec k)=\widehat\sigma_{k\phik}(\vec k)=0$ for any $\vec k\ne\vec 0$. In the RT representation [see Eq.~(\ref{eq:stress:rst})], it becomes:
\begin{equation}\label{eq:me}
\forall\vec k\ne\vec 0\qquad
\left\{
\begin{aligned}
\widehat\sigma_{2}^{\uvec k}(\vec k)&=\frac{1}{\sqrt{2}}\,\widehat\sigma_{1}^{\uvec k}(\vec k)\\
\widehat\sigma_{3}^{\uvec k}(\vec k)&=\widehat\sigma_{4}^{\uvec k}(\vec k)=0
\end{aligned}
\right.
\end{equation}
Mechanical balance alone therefore implies that the stress autocorrelation matrix $\uttilde{\widehat C}_{\vec k}$ is of the form:
\begin{equation}
  \label{eq:mechanical:C}
\begin{split}
&\forall\vec k\ne\vec 0\qquad\mathring{\uttilde{\widehat{C}}}(\vec k)=\\[2mm]
&\left(
  \begin{matrix}
\ \mathring{\widehat{C}}^{(1)}(\vec k)& \frac{1}{\sqrt{2}}\,\mathring{\widehat{C}}^{(1)}(\vec k)& 0\quad & 0 \quad &\ \mathring{\widehat{C}}^{(5)}(\vec k)&\ \mathring{\widehat{C}}^{(6)}(\vec k)\\[1mm]
\ \frac{1}{\sqrt{2}}\,\mathring{\widehat{C}}^{(1)}(\vec k)& \frac{1}{2}\,\mathring{\widehat{C}}^{(1)}(\vec k)&0\quad & 0 \quad &\frac{1}{\sqrt{2}}\,\mathring{\widehat{C}}^{(5)}(\vec k)&\frac{1}{\sqrt{2}}\,\mathring{\widehat{C}}^{(6)}(\vec k)\\[1.7mm]
\ 0& \ 0& 0\quad &0 \quad &\ 0&\ 0\\[1.7mm]
\ 0& \ 0& 0\quad & 0 \quad &\ 0&\ 0\\[1.7mm]
\ \mathring{\widehat{C}}^{(5)}(\vec k)&\frac{1}{\sqrt{2}}\,\mathring{\widehat{C}}^{(5)}(\vec k)&0\quad & 0 \quad &\ \mathring{\widehat{C}}^{(2)}(\vec k)&\ \mathring{\widehat{C}}^{(4)}(\vec k)\\[1.7mm]
\ \mathring{\widehat{C}}^{(6)}(\vec k)&\frac{1}{\sqrt{2}}\,\mathring{\widehat{C}}^{(6)}(\vec k)&0\quad & 0 \quad &\ \mathring{\widehat{C}}^{(4)}(\vec k)&\ \mathring{\widehat{C}}^{(3)}(\vec k)
\end{matrix}
\right)
\end{split}
\end{equation}
with six scalar, not necessarily isotropic, functions $\mathring{\widehat{C}}^{(i)}$, $i=1,\ldots,6$.

It now appears that, when both mechanical balance [Eq.~(\ref{eq:mechanical:C})] and material isotropy [Eq.~(\ref{eq:constraint:k})] hold, the inherent stress correlation $\mathring{\uttilde{\widehat{C}}}({\vec k})$ must present the following remarkably simple structure:
\begin{equation}
\label{eq:all}
\begin{split}
&\forall\vec k\ne\vec 0\qquad\\
&\mathring{\uttilde{\widehat{C}}}({\vec k})=\left(
  \begin{matrix}
 \mathring{\widehat{C}}_1(k)&\frac{1}{\sqrt{2}}\,\mathring{\widehat{C}}_1(k) &  0&\quad 0  &\ 0 &\ 0\\[1.7mm]
\!\!\frac{1}{\sqrt{2}}\,\mathring{\widehat{C}}_1(k)\!\!& \ \frac{1}{2}\,\mathring{\widehat{C}}_1(k)& 0 &\quad 0 &\ 0&\ 0\\[1.7mm]
 0&  0&  0 &\quad  0 &\ 0&\ 0\\[1.7mm]
 0&  0& 0 &\quad  0 &\ 0&\ 0\\[1.7mm]
 0&  0& 0 &\quad  0 &\ \mathring{\widehat{C}}_5(k)&\ 0\\[1.7mm]
 0&  0& 0 &\quad  0 &\ 0&\ \mathring{\widehat{C}}_5(k)
  \end{matrix}
\right)
\end{split}
\end{equation}
Let us emphasize that we are dealing here with a fourth order tensor field, i.e. an object that---after eliminating the trivial matrix symmetry---comprises a priori 21 scalar fields. We have just demonstrated that the combination of mechanical balance and material isotropy reduces it to take a form that only involves two scalar functions $\mathring{\widehat{C}}_1$ and $\mathring{\widehat{C}}_5$ of the magnitude $k$, which is a considerable simplification. Moreover, these two functions are of course non-negative.


Thanks to our previous identification of the isotropic parts of the stress autocorrelation, it is straightforward to relate $\mathring{\widehat{C}}_1$ and $\mathring{\widehat{C}}_5$ to the two functions ${C}_0$ and ${C}_0'$ defined in Eq.~(\ref{eq:trace:cs}). Indeed, using Eq.~(\ref{eq:trace:rs:Fourier}), we may compute the coefficients $\widehat{C}_0$ and $\widehat{C}_0'$ of $\uttilde{\widehat{C}}^{\rm iso}$ as:
\begin{equation}\label{eq:C:0:inverse}
\begin{split}
\widehat{C}_0(k)&=\mathring{\widehat{C}}_1(k)\\
\widehat{C}_0'(k)&=\frac{1}{10}\,\left(\mathring{\widehat{C}}_1(k)+4\,\mathring{\widehat{C}}_5(k)\right)
\end{split}
\end{equation}
As already said [Eq.~(\ref{eq:isoft})] these functions are just (as the notation suggests) the (3D) Fourier transforms of the radial functions ${C}_0$ and ${C}_0'$ (resp.).
This system of equations is easily inverted to write $\mathring{\widehat{C}}_1$ and $\mathring{\widehat{C}}_5$ in terms of $\widehat{C}_0$ and $\widehat{C}_0'$. We have thus established that the pressure autocorrelation $C_0$ and the isotropic part $C_0'$ of the autocorrelation of stress deviators fix the complete form of the stress autocorrelation in an infinite medium.


\subsubsection{Real space form}
\label{sec:real}

The real space form of the stress autocorrelation is already known to be of the form~(\ref{eq:constraint:C}), thanks to material isotropy,  i.e. to be fixed by the five functions $\mathring{C}_{i}$. The calculation of these functions from the Fourier space expression~(\ref{eq:all}), which requires to proceed through the relations expressed diagramatically in~(\ref{diag}), is performed in  Appendix~\ref{sec:compute} and yields:
\begin{equation}\label{eq:C:inverse}
\begin{split}
\mathring{C}_{1}(r)&=\mathring{C}_1^{(0)}\\
\mathring{C}_{2}(r)&=-\frac{\sqrt{2}}{2}\,\mathring{C}_1^{(2)}\\
\mathring{C}_{3}(r)&=\frac{\mathring{C}_1^{(0)}+4\mathring{C}_5^{(0)}}{10}-\frac{\mathring{C}_1^{(2)}-4\mathring{C}_5^{(2)}}{7}+\frac{9\mathring{C}_1^{(4)}+6\mathring{C}_5^{(4)}}{35}\\
\mathring{C}_{4}(r)&=\frac{\mathring{C}_1^{(0)}+4\mathring{C}_5^{(0)}}{10}-\frac{\mathring{C}_1^{(2)}-4\mathring{C}_5^{(2)}}{14}-\frac{6\mathring{C}_1^{(4)}+4\mathring{C}_5^{(4)}}{35}\\
\mathring{C}_{5}(r)&=\frac{\mathring{C}_1^{(0)}+4\mathring{C}_5^{(0)}}{10}+\frac{\mathring{C}_1^{(2)}-4\mathring{C}_5^{(2)}}{7}+\frac{3\mathring{C}_1^{(4)}+2\mathring{C}_5^{(4)}}{70}
\end{split}
\end{equation}
with the following transform:
\begin{equation}\label{eq:Cm}
\mathring{C}^{(m)}(r)=(2\pi)^{-3/2}\,\int_0^\infty\d k\ k^2\,\mathring{\widehat{C}}(k)\,\frac{J_{m+\frac{1}{2}}(kr)}{\sqrt{kr}}
\end{equation}
where $\mathring{\widehat{C}}$ stands for $\mathring{\widehat{C}}_1$ or $\mathring{\widehat{C}}_5$, and with $J_m$ the Bessel function of the first kind.
We have chosen to write the autocorrelation components using transforms of $\mathring{\widehat{C}}_1$ or $\mathring{\widehat{C}}_5$. We could identically have reorganized the terms to involve transforms of the functions $\widehat{C}_0=\mathring{\widehat{C}}_1$ and $\widehat{C}_0'$; the corresponding expressions are provided at the end of Appendix~\ref{sec:compute}; there is no difference in proceeding either way.

Using either Equation~(\ref{eq:C:0:inverse}) or~(\ref{eq:trace:rs}), we recognize the isotropic contributions to $\uttilde{C}$:
\begin{compactitem}
\item the pressure autocorrelation $\mathring{C}_1(r)=\mathring{C}_1^{(0)}\equiv C_0$;
\item the first term, $C_0'=\frac{1}{10}(\mathring{C}_1^{(0)}+4\mathring{C}_5^{(0)})$, in $\mathring{C}_3$, $\mathring{C}_4$, and $\mathring{C}_5$.
\end{compactitem}
All other terms contribute to the anisotropic part $\mathring{\uttilde{C}}^{\rm ani}=\mathring{\uttilde{C}}-\mathring{\uttilde{C}}^{\rm iso}$. Note that isotropic terms only involve the functions $\mathring{C}_1^{(m)}(r)$ and $\mathring{C}_5^{(m)}(r)$ with $m=0$, while anisotropic terms only involve $m=2$ or $4$ transforms.

It is not obvious that the integrals appearing in equation~(\ref{eq:Cm}) are always well-defined. But in the case $m=0$, Eq.~(\ref{eq:Cm}) it is just the inverse Fourier transform in 3D. The well-definedness of $m\ne0$ transforms starts to emerge once we realize they also are inverse Fourier transforms, yet in higher-dimensional spaces. Let us recall, indeed, that in $\mathbb{R}^d$, a radial scalar function $f(r)$ is related to its Fourier transform by:
\begin{equation}
f(r)=(2\pi)^{-d/2}\,\int_0^\infty\d k\ \frac{k^{d/2}}{r^{d/2-1}}\,\widehat{f}(k)\,J_{\frac{d}{2}-1}(kr)
\end{equation}
Comparing this equation with~(\ref{eq:Cm}) we find that for all $m$ (which includes $m=0$ as a particular case):
\begin{equation}\label{eq:inverse:d}
\mathring{C}^{(m)}(\|\vec r\|)=(2\pi)^m\,r^m\,\mathcal{F}_{2m+3}^{-1}\left[\frac{\mathring{\widehat{C}}(\|\vec k\|)}{\|\vec k\|^m}\right]
\end{equation}
in which $\mathring{C}^{(m)}(\|\vec r\|)$ and $\mathring{\widehat{C}}(\|\vec k\|)$ are understood as radial functions of (resp.) the vectors $\vec r,\vec k\in\mathbb{R}^{d}$, with $d=2m+3$, and where $\mathcal{F}_{d}$ is the Fourier transform in $\mathbb{R}^d$. Shortly, we will see more precisely under what conditions these transforms are well-defined.

\subsubsection{Real-space asymptotic decay}
\label{sec:decay}

Let $\mathring{\widehat{C}}$ denote either $\mathring{\widehat{C}}_1$ or $\mathring{\widehat{C}}_5$. As we are interested in characterizing the dominant contributions to the decay of correlations in real-space, we will set aside any issue related to possible small scale singularities and will consider, possibly after the introduction of an arbitrary short-scale regularization, that $\mathring{\widehat{C}}$ has all needed regularity property for $k\ne0$ and decays as rapidly as needed in the $k\to\infty$ limit. The real-space decay of the functions $\mathring{C}^{(m)}$ for $m=0$, 2, and 4 is thus controlled by the behavior of $\mathring{\widehat{C}}$ near $k=0$.

As explained previously, we are mainly concerned by glasses, in which sphere-averaged stress fluctuations abide by the normal decay, i.e. for which $\mathring{\widehat{C}}(\vec 0)$ is a constant. We will expound this case before commenting briefly on other eventualities.\\

The following discussion will make repeated use of the well-known property that for any $d+s>0$, provided $s\ne 0, 2, 4,\ldots$, the inverse Fourier transform of $k^s$, which is defined in the sense of tempered distributions in $\mathbb{R}^d$, is:
\begin{equation}\label{eq:inverse:explicit}
  \mathcal{F}_{d}^{-1}\left[k^s\right]=\frac{c_{d,s}}{r^{d+s}}
\end{equation}
with the constant
\begin{equation}
c_{d,s}=\frac{2^s}{\pi^{\frac{d}{2}}}\,\frac{\Gamma\left(\frac{d+s}{2}\right)}{\Gamma\left(-\frac{s}{2}\right)}
\end{equation}
It is important that this relation applies, in particular, to all values of $s$ on the interval $0>s>-d$.\\

We first examine the real-space decay of $\mathring{C}^{(0)}$, i.e. taking $m=0$, $d=3$. Two cases must be distinguished:
\begin{compactenum}
\item[A.] $\mathring{\widehat{C}}(\vec k)$ is analytic at the origin, i.e. all the exponents $s$ appearing in its small $k$ expansion, $\mathring{\widehat{C}}(k)=\mathring{\widehat{C}}(0)+A\,k^s+\cdots$ are non-negative even integers. In that case, $\mathring{\widehat{C}}(\|\vec k\|)$ is a rapidly decaying function and so is its Fourier inverse $\mathring{C}^{(0)}(\vec r)$, which hence essentially decays exponentially at large distances. This guarantees the existence of a characteristic decay length for the pressure autocorrelation.
\item[B.] $\mathring{\widehat{C}}(k)$ is non-regular at the origin, for example, it is of the form $\mathring{\widehat{C}}(k)=\mathring{\widehat{C}}(0)+A\,k^s+\cdots$ with some $s>0$ that is not an even integer. As shown by expression~(\ref{eq:inverse:explicit}) the inverse Fourier transform of $A\,k^s$ is $A\,c_{3,s}/r^{3+s}$. It decays as a power law in space, yet with an exponent $3+s$, which is larger than $3$. (Note that the real-space decay $A'/r^{3+s}$ with $s=2,4,\ldots$ does not correspond to terms of the power-law form $A\,k^s$ in $\mathring{\widehat{C}}(k)$; they correspond instead to terms $\propto -k^s\ln k$.)
\end{compactenum}\mbox{}\\

Coming now to the functions $\mathring{C}^{(m)}$ with $m=2$ and 4, let us rewrite Eq.~(\ref{eq:inverse:explicit}) for the tempered distribution $k^{-m}$, i.e. in the specific case $s=-m$, noting that we are only concerned by $d=2m+3$ and hence $0<m<d$:
\begin{equation}
  \mathcal{F}_{d}^{-1}\left[\frac{1}{k^m}\right]=\frac{c_{d,-m}}{r^{d-m}}
\end{equation}
Equation~(\ref{eq:inverse:d}) shows that the function $\mathring{C}^{(m)}/r^m$ is the convolution of $\mathcal{F}_{d}^{-1}\left[\mathring{\widehat{C}}(\|\vec k\|)\right]$ by the kernel $r^{-(d-m)}$. Such convolutions, with $0<m<d$, are called Riesz potentials~\cite{Riesz1949,Landkof1972}. They are defined provided the convolved function decays sufficiently rapidly at infinity, a requirement that always applies in our case~\footnote{Riesz potentials are well-defined, in arbitrary $d$, provided the convolved functions belongs to the functional space $L^q(\mathbb{R}^d)$ for some $1\le q<d/m$. This tight requirement always applies in our case, since the assumed regularity at $\vec k\ne\vec 0$ and the finiteness of {$\mathring{\widehat{C}}(\vec k=\vec 0)$} guarantees that, in any $d$, the radial function {$\mathring{\widehat{C}}(\|\vec k\|)$} is in $L^p(\mathbb{R}^d)$ for any $1\le p\le\infty$; thus, its inverse Fourier transform {$\mathcal{F}_{d}^{-1}\left[\mathring{\widehat{C}}(\vec k)\right]$} is in $L^q(\mathbb{R}^d)$ for any $2\le q\le\infty$. The property $d/m=2+3/m>2$ finally insures that we can find $1\le q<d/m$ so that the corresponding Riesz potential, hence the transformation~(\protect\ref{eq:Cm}), is always well-defined.}.

The real-space form of $\mathring{C}^{(m)}$ for $m=2$, 4, corresponding to the approximation $\mathring{\widehat{C}}(\vec k)\simeq\mathring{\widehat{C}}(\vec 0)+\cdots$ is:
\begin{equation}
\frac{\mathring{C}^{(m)}(r)}{(2\pi)^m}= \mathring{\widehat{C}}(0)\frac{c_{2m+3,-m}}{r^3}+\cdots
\end{equation}
It is instructive to examine how the behavior of $\mathring{\widehat{C}}(k)-\mathring{\widehat{C}}(0)$ in the vicinity of $k=0$ contributes to subdominant terms in the large-$r$ expansion of $\mathring{C}^{(m)}$. So we write $\mathring{\widehat{C}}(k)=\mathring{\widehat{C}}(0)+A\,k^s+\cdots$ and distinguish two cases:
\begin{compactenum}
\item[A'.] If $s-m\ge0$ is a non-negative even integer, the corresponding real-space contribution is a rapidly decaying function.
\item[B'.] Otherwise, which includes both the case $s<m$ when $A\,k^s$ corresponds to another Riesz potential and $s>m$ when it is an increasing function of $k$, we find:
\begin{equation}\label{eq:decay:s}
\frac{\mathring{C}^{(m)}(r)}{(2\pi)^m}= \mathring{\widehat{C}}(0)\frac{c_{2m+3,-m}}{r^3}+A\,\frac{c_{2m+3,s-m}}{r^{3+s}}+\cdots
\end{equation}
\end{compactenum}

Clearly, for $m=2$ and, 4 the leading far-field contribution to the long-range decay of the functions $\mathring{C}^{(m)}$'s is always:
\begin{equation}\label{eq:decay:3}
\mathring{C}^{(m)}(r)\simeq A_m\frac{\mathring{\widehat{C}}(0)}{r^3}+\cdots
\end{equation}
with
\begin{equation}
A_m\equiv (2\pi)^m\,c_{2m+3,-m}=\frac{1}{\pi^{3/2}}\,\frac{\Gamma\left(\frac{m+3}{2}\right)}{\Gamma\left(\frac{m}{2}\right)}
\end{equation}
the two relevant values of which are: $A_2=3/(4\pi)$ and $A_4=15/(8\pi)$.
This result should be contrasted with the analysis of $\mathring{C}^{(0)}$, where we found that the slowest ($1/r^{s+3}$) contribution to the real-space decay is \emph{not} obtained at zeroth order, but from the presence of $A\,k^s$ terms in the low-$k$ expansion. In the case $m=0$, when we approximate $\mathring{\widehat{C}}(\vec k)\simeq\mathring{\widehat{C}}(\vec 0)+\cdots$, we replace the function $\mathring{\widehat{C}}/k^{m}$, appearing at the right hand side of Eq.~(\ref{eq:inverse:d}) by an analytical function that corresponds to a rapidly decaying function in real space; in this $m=0$ case, the lack of regularity at $k=0$ hence may only come from higher order terms in the expansion. In sharp constrast, for $m=2$ or 4, the approximation $\mathring{\widehat{C}}(\vec k)\simeq\mathring{\widehat{C}}(\vec 0)+\cdots$ always captures the most singular contribution to $\mathring{\widehat{C}}/k^{m}$, and thus the leading order term in the real-space decay.\\

Let us now put all the above results together and turn to examining the asympotic behavior of $\uttilde{C}$, the real-space stress autocorrelation function, the RT form of which is specified by Eq.~(\ref{eq:C:inverse}). The functions $\mathring{C}_1^{(0)}$ and $\mathring{C}_5^{(0)}$ only contribute to the isotropic part ${\uttilde{C}}^{\rm iso}(r)$ of $\uttilde{C}$. The above analysis has shown that the field ${\uttilde{C}}^{\rm iso}(r)$ may either decay rapidly or as slowly as $1/r^{3+s}$ with some $s>0$.

The anisotropic part $\mathring{\uttilde{C}}^{\rm ani}(r)$ only involves the functions $\mathring{C}_1^{(m)}(r)$ and $\mathring{C}_5^{(m)}(r)$ with $m=2$ or 4. To examine its decay we still have to report Eq.~(\ref{eq:decay:3}) in Eq.~(\ref{eq:C:inverse}), which yields:
\begin{equation}\label{eq:ani:decay}
\begin{split}
\mathring{C}_{1}^{\rm ani}(r)&=0\\
\mathring{C}_{2}^{\rm ani}(r)&=-\frac{3\sqrt{2}}{8\pi}\,\mathring{\widehat{C}}_1(0)\ \frac{1}{r^3}+\cdots\\
\mathring{C}_{3}^{\rm ani}(r)&=\left(\frac{3}{8\pi}\mathring{\widehat{C}}_1(0)+\frac{3}{4\pi}\mathring{\widehat{C}}_5(0)\right)\ \frac{1}{r^3}+\cdots\\
\mathring{C}_{4}^{\rm ani}(r)&=-\frac{3}{8\pi}\mathring{\widehat{C}}_1(0)\ \frac{1}{r^3}+\cdots\\
\mathring{C}_{5}^{\rm ani}(r)&=\left(\frac{3}{16\pi}\mathring{\widehat{C}}_1(0)-\frac{3}{8\pi}\mathring{\widehat{C}}_5(0)\right)\ \frac{1}{r^3}+\cdots
\end{split}
\end{equation}
It must be recalled that both $\mathring{\widehat{C}}_1(0)$ and $\mathring{\widehat{C}}_5(0)$ are positive. Therefore, the tails of $\mathring{C}_{2}^{\rm ani}$ and $\mathring{C}_{4}^{\rm ani}$ are systematically negative, while that of $\mathring{C}_{3}^{\rm ani}(r)$ is always positive; the $1/r^3$ term in $\mathring{C}_{5}^{\rm ani}(r)$ may have either sign, and may vanish if $\mathring{\widehat{C}}_5(0)=\mathring{\widehat{C}}_1(0)/2$. It is striking that the decay of the anisotropic part is thus systematically a $1/r^3$ power law, irrespective of the real-space decay of the isotropic components.\\

To close this discussion, let us briefly examine the cases when the function $\mathring{\widehat{C}}=\mathring{\widehat{C}}_1$ or $\mathring{\widehat{C}}_5$ either diverges or vanishes at $k=0$.
\begin{compactitem}
\item[A''.] If $\mathring{\widehat{C}}\simeq A/k^{s}$ around $k=0$, with $0<s<3$, it is easy to see from our previous analysis that all the functions $\mathring{C}^{(m)}/r^m$ are Riesz potentials, since $0<m+s<2m+3$, so that all the $\mathring{C}^{(m)}$'s decay as $1/r^{3-s}$ in real space.
\item[B''.]
If $\mathring{\widehat{C}}(0)=0$, we must consider three subcases:
\begin{compactitem}
\item[1.] if $\mathring{\widehat{C}}(k)\simeq Ak^2$ (assuming regularity of all other contributions), then $\mathring{C}^{(0)}$ and $\mathring{C}^{(2)}$ are rapidly decaying in real space, while $\mathring{C}^{(4)}$ decays as $1/r^5$.
\item[2.] if $\mathring{\widehat{C}}(k)\simeq Ak^s$ with $s$ an even integer $\ge4$, then the whole stress autocorrelation function is rapidly decaying in real space.
  \item[3.] if $\mathring{\widehat{C}}(k)\simeq Ak^s$ where $s>0$ is not an even integer, all the functions $\mathring{C}^{(m)}$ for $m=0$, 2, and 4, hence the whole stress autocorrelation function, decay as $1/r^{3+s}$ in real space.
\end{compactitem}
\end{compactitem}
This enumeration illustrates that in most cases, except the hypothetical case B''.2. above, the stress autocorrelation presents power law tails. Yet, the only case when a component of the stress autocorrelation decays as $1/r^3$ is when $\mathring{\widehat{C}}(0)$ is finite---this result may also be read from the inversion of Eq.~(\ref{eq:inverse:d}). The $1/r^3$ decay of the anisotopic components of the stress autocorrelation therefore appears to be characteristic of isotropic solids that present normal fluctuations of the sphere-averaged stress.


\section{Conclusion}

In this article, we have demonstrated that the presence of long-range $1/r^3$ correlation tails in the stress autocorrelation of 3D glass is an analytical consequence of material isotropy and mechanical balance, for systems in which the fluctuations of the sphere-averaged stress present the normal inverse volume decay. This result has been obtained without any material-specific assumption other than concerning the scaling of stress fluctuations.

Our paper also contains a number of results that are much more general than the case of glassy solids that motivated our initial interest in the topic of stress correlations.
Our formalism and discussion of material isotropy (most of Sec.~\ref{sec:materialisotropy}), which is taken from~\cite{Lemaitre2015}, is independent of mechanical balance a may be used to analyze, e.g. stress in parent liquid configurations. Our identification of the spatially isotropic contributions to the stress autocorrelation matrix (Sec.~\ref{sec:isotropy}) and its relation to the sphere-averaged stress fluctuations also applies to any isotropic system. Regarding this matter, let us emphasize that, while it is classical that the pressure autocorrelation is radial in an isotropic medium, the existence and importance of the second isotropic contribution seem to have been completely overlooked in the literature. It is classical, of course, that the trace of a matrix is rotation-invariant, but it seems it was not recognized previously that the function $C_0'$ is independent of $C_0$ in 3D isotropic media, and shares with the fluctuations of the sphere-averaged deviatoric stresses the same relation as $C_0$ to the fluctuations of the sphere-averaged pressure.

The main results of this article were obtained in Section~\ref{sec:compounding} where we have derived the general form of stress correlation for isotropic and mechanicaly balanced systems [Eq.~(\ref{eq:all}) in Fourier space and Eqs.~(\ref{eq:C:inverse}) and~(\ref{eq:inverse:d}) in real space]. These expressions hold independently of the behavior of stress fluctuations in the large averaging domain limit. By analysing the three possible cases for this behavior, we have emphasized that mechanical balance and material isotropy alone do not guarantee the presence of $1/r^3$ correlation tails. Yet, the specific analytic relation they enforce between isotropic and anisotropic terms guarantees that the anisotropic contributions decay as $1/r^3$ as soon as stress fluctuations are normal. Long-range, $1/r^3$ anisotropic stress correlations hence exist in all isotropic solids that present normal of stress fluctuations.

Let us emphasize that local stress fluctuations may not be normal in all ensembles of isotropic solids: a counter-example exist, as non-self-averaging macroscopic stress fluctuations were recently reported~\cite{WuKarimiMaloneyTeitel2017}, yet in systems produced after direct quenches from random configurations, without any equilibration. Existing data on stress fluctuations in inherent states of equilibrated liquids~\cite{AbrahamHarrowell2012,Lemaitre2015} or in granular systems~\cite{HenkesChakraborty2009}, however, show normal stress fluctuations. As we already argued, this normal behavior should be the rule in glasses because these systems should present a well-defined stress state in the thermodynamic limit. For this reason, it seems utterly reasonable to expect that stress fluctuation are normal in glasses and thus to conclude for the generic presence of $1/r^3$ stress correlation tails in these systems.



It is left to future works to assess what role these stress correlations may play in phenomena such as sound damping or supercooled liquid relaxation. In this regard, the tightness of the mathematical constraints laid upon the stress autocorrelation are, in a certain sense, rather perplexing. Indeed, if stress correlations have a role in e.g. relaxation, there should be some qualitative feature that distinguishes stress correlations in fragile and strong glasses. Our result implies that all glasses carry $1/r^3$ power-law correlations that are essentially identical. The interest of our analysis is also that we can reduce the problem of characterizing the full stress autocorrelation of glasses to the study of just two scalar and radial functions $C_0$ and $C_0'$. But we are now left wondering what distinctive features of the pressure autocorrelation and its deviatoric counterpart would matter to discriminate different types of glasses.

\thanks

Illuminating discussions with Christiane Caroli, Fr\'ed\'eric Legoll, and Antoine Levitt are gratefully acknowledged.

\appendix

\section{Local stress fluctuations}
\label{sec:integrals}

We here recall and adapt to our case some results from Refs~\cite{TorquatoStillinger2003,Torquato2016,Torquato2018}. The scaled intersection volume function defined in Eq.~(\ref{def:alpha}) reads explicitly~\cite{TorquatoStillinger2003,Torquato2018}:
\begin{equation}\label{eq:alpha}
 \alpha(\vec r; R)=\left\{
\begin{aligned}
  & 1-\frac{3}{4}\frac{r}{R}+\frac{1}{16}\,\left(\frac{r}{R}\right)^3\qquad \textrm{if}\ \|\vec r\|<2R\\
  &0\qquad \textrm{otherwise}
\end{aligned}\right.
\end{equation}
It is a radial function that smoothly decreases from 1 to 0 as $r$ increases from 0 to $2R$. Its Fourier transform~\cite{Torquato2018} reads
\begin{equation}
\widehat\alpha(\vec k;R)= {6\pi^2}\frac{\left[J_{3/2}(kR)\right]^2}{k^3}
\end{equation}
with $J_m$ the Bessel function of the first kind.
It is a non-negative decaying oscillating function of $\|\vec k\|$, which integrates to unity
\begin{equation}
\int\frac{\d^3\vec k}{(2\pi)^3}\ \widehat\alpha(\vec k;R)=1
\end{equation}
so that it converges to $(2\pi)^3\,\delta(\vec k)$ when $R\to\infty$~\cite{TorquatoStillinger2003,Torquato2018}.

Using these definitions, Eq.~(\ref{eq:JR}) may be written~\cite{TorquatoStillinger2003,Torquato2016,Torquato2018}:
\begin{equation}\label{eq:J:alpha}
\uttilde J(R)
=\frac{1}{\Omega_R}\ \int\d^3\vec r\ \alpha(\vec r; R)\,\uttilde C^{\rm iso}\left(\vec r\right)
\end{equation}
or equivalently:
\begin{equation}
\uttilde{J}(R)=\frac{1}{\Omega_R}\ \int\,\frac{\d^3\vec k}{(2\pi)^3}\,\widehat\alpha(\vec k;R)\,\widehat{\uttilde C}^{\rm iso}\left(\vec k\right)
\end{equation}

Using Eq.~(\ref{eq:J:alpha}) and~(\ref{eq:alpha}), the general $R$-dependence of the sphere-averaged stress fluctuations is easily deduced using the real space form of $\alpha$ [see Ref.~\cite{Torquato2018} and appendix~\ref{sec:integrals}]:
\begin{equation}
\begin{split}
  \uttilde{J}(R)&=\frac{1}{\Omega_R}\,\int_{r<R}\d^3\vec r\,\uttilde C^{\rm iso}\left(\vec r\right)\\
  &-\frac{3}{4R\Omega_R}\,\int_{r<R}\d^3\vec r\,r\,\uttilde C^{\rm iso}\left(\vec r\right)\\
  &+\frac{\pi}{12\Omega_R^2}\,\int_{r<R}\d^3\vec r\,r^3\,\uttilde C^{\rm iso}\left(\vec r\right)
\end{split}
\end{equation}
which shows that $\uttilde C^{\rm iso}$ completely fixes the large $R$ behavior of $\uttilde{J}$.

\section{Infinite medium expression in real space}
\label{sec:compute}

Equation~(\ref{eq:all}) may be recast as:
\begin{equation}
\forall\vec k\ne\vec 0\qquad{\mathring{\uttilde{\widehat{C}}}}({\vec k}) = \mathring{\widehat{C}}_1(k)\mathring{\uttilde{A}}+\mathring{\widehat{C}}_5(k)\mathring{\uttilde{B}}
\end{equation}
which defines the two constant matrices $\mathring{\uttilde{A}}$ and $\mathring{\uttilde{B}}$. The corresponding CT form reads:
\begin{equation}
\forall\vec k\ne\vec 0\qquad{\uttilde{\widehat{C}}}({\vec k}) = \mathring{\widehat{C}}_1(k)\uttilde{A}(\uvec k)+\mathring{\widehat{C}}_5(k)\uttilde{B}(\uvec k)
\end{equation}
where
\begin{equation}
\begin{split}
\uttilde{A}(\uvec k)&=(\tttt{D}^{\uvec k})^T\cdot\mathring{\uttilde{A}}\cdot\tttt{D}^{\uvec k}\\
\uttilde{B}(\uvec k)&=(\tttt{D}^{\uvec k})^T\cdot\mathring{\uttilde{B}}\cdot\tttt{D}^{\uvec k}
\end{split}
\end{equation}
only depend on the direction $\uvec k$ of vector $\vec k$, but not on its amplitude. Hopefully, we do not need to compute the complete form of these matrices.

Our goal is to calculate the real-space RT form $\mathring{\uttilde{C}}(r)$ which, in view of Eq.~(\ref{eq:z}), may be obtained as:
\begin{equation}
\mathring{\uttilde{C}}(r)=\frac{1}{(2\pi)^3}\int\d\vec k\ e^{ir\,\vec k\cdot\vec e_z}\ {\uttilde{\widehat{C}}}({\vec k})
\end{equation}
Introducing the integrals in
\begin{equation}\label{eq:IAB}
\begin{split}
\uttilde{I}^A(z)&=\frac{1}{4\pi}\,\int\d\uvec k\ e^{iz\,\uvec k\cdot\vec e_z}\ \uttilde{A}(\uvec k)\\
\uttilde{I}^B(z)&=\frac{1}{4\pi}\,\int\d\uvec k\ e^{iz\,\uvec k\cdot\vec e_z}\ \uttilde{B}(\uvec k)
\end{split}
\end{equation}
it comes:
\begin{equation}\label{eq:RT:all}
  \mathring{\uttilde{C}}(r)=\frac{1}{2\pi^2}\int\d k\ k^2\,\left(\mathring{\widehat{C}}_1(k)\,I^A(kr)+\mathring{\widehat{C}}_5(k)\,I^B(kr)\right)
\end{equation}
Since the real space RT autocorrelation $\mathring{\uttilde{C}}(r)$ is necessarily of the form~(\ref{eq:constraint:C}) for any value of the functions $\mathring{\widehat{C}}_1$ and $\mathring{\widehat{C}}_5$, the fields $\uttilde{I}^A$ and $\uttilde{I}^B$ are also of the form~(\ref{eq:constraint:C}), hence have only five non-zero coefficients, $I_i^A$ and $I_i^B$ (resp.), with $i=1,\ldots,5$. We only need to compute these from the corresponding matrix elements of $\uttilde{A}(\uvec k)$ and $\uttilde{B}(\uvec k)$ [these latter two matrices, however, have many more non-zero components, since they define the (Fourier) CT form and do not comply with~(\ref{eq:constraint:C})].

Using the spherical coordinates $\vec k=(k,\thetak,\phik)$, and the expression of $\tttt{D}^{\uvec k}$ from~\cite{Lemaitre2015}, we find:
\begin{equation}
\begin{split}
A_{11}&=1\\
A_{12}=A_{21}&=\frac{\sqrt{2}}{2}-\frac{3\sqrt{2}}{4}\sin^2\thetak\\
A_{22}&=\frac{1}{2}-\frac{3}{2}\sin^2\thetak+\frac{9}{8}\sin^4\thetak\\
A_{33}&=\frac{3}{4}\left(\sin^2\thetak-\sin^4\thetak\right)\left(1-\cos2\phik\right)\\
A_{44}&=\frac{3}{4}\left(\sin^2\thetak-\sin^4\thetak\right)\left(1+\cos2\phik\right)\\
A_{55}&=\frac{3}{16}\sin^4\thetak\left(1-\cos4\phik\right)\\
A_{66}&=\frac{3}{16}\sin^4\thetak\left(1+\cos4\phik\right)\\
\end{split}
\end{equation}
and
\begin{equation}
\begin{split}
B_{11}&=0\\
B_{12}=B_{21}&=0\\
B_{22}&=\frac{3}{4}\sin^4\thetak\\
B_{33}&=\sin^2\thetak-\frac{1}{2}\sin^4\thetak+\frac{1}{2}\sin^4\thetak\cos2\phik\\
B_{44}&=\sin^2\thetak-\frac{1}{2}\sin^4\thetak-\frac{1}{2}\sin^4\thetak\cos2\phik\\
B_{55}&=1-\sin^2\thetak+\frac{1}{8}\sin^4\thetak-\frac{1}{8}\sin^4\thetak\cos4\phik\\
B_{66}&=1-\sin^2\thetak+\frac{1}{8}\sin^4\thetak+\frac{1}{8}\sin^4\thetak\cos4\phik
\end{split}
\end{equation}

Now we use Eq.~(\ref{eq:IAB}) with $\int\d\uvec k=\int_0^\pi\d\thetak\int_0^{2\pi}\d\phik\,\sin(\thetak)$. When integrating the above functions over $\phik$, all the terms proportional to $\cos n\phik$ vanish. The non-zero coefficients of $\uttilde{I}^A$ (and likewise of $\uttilde{I}^B$) are hence of the form
\begin{equation}
I^A_{i}(z)=\frac{1}{2}\,\int_0^\pi\d\thetak\ e^{iz\cos\thetak}P_{i}^A(\sin\thetak)
\end{equation}
with $P_{i}$ odd polynomials. These integrals are calculated using
\begin{equation}
\begin{split}
\frac{1}{2}\,\int_0^\pi\d\thetak\,e^{iz\cos\thetak}\sin^{2n+1}\thetak
&=2^{n}n!\, z^{-n} j_{n}(z)
\end{split}
\end{equation}
which holds for any non-negative integer $n$, leading to:
\begin{equation}
\begin{split}
I^A_{1}(z)&=j_0(z)\\
I^A_{2}(z)&=\frac{\sqrt{2}}{2}\,j_0(z)-\frac{3\sqrt{2}}{2}\frac{j_1(z)}{z}\\
I^A_{3}(z)&=\frac{1}{2}\,j_0(z)-3\,\frac{j_1(z)}{z}+9\,\frac{j_2(z)}{z^2}\\
I^A_{4}(z)&=\frac{3}{2}\,\frac{j_1(z)}{z}-6\,\frac{j_2(z)}{z^2}\\
I^A_{5}(z)&=\frac{3}{2}\,\frac{j_2(z)}{z^2}\\
\end{split}
\end{equation}
and
\begin{equation}
\begin{split}
I^B_{1}(z)&=0\\
I^B_{2}(z)&=0\\
I^B_{3}(z)&=6\,\frac{j_2(z)}{z^2}\\
I^B_{4}(z)&=2\,\frac{j_1(z)}{z}-4\,\frac{j_2(z)}{z^2}\\
I^B_{5}(z)&=j_0(z)-2\,\frac{j_1(z)}{z}+\frac{j_2(z)}{z^2}
\end{split}
\end{equation}
We next use the recurrence relation
\begin{equation}
\frac{j_{n}(z)}{z}=\frac{j_{n-1}(z)+j_{n+1}(z)}{2n+1}
\end{equation}
to find
\begin{equation}
\begin{split}
\frac{j_1}{z}&=\frac{j_0+j_2}{3}\\
\frac{j_2}{z^2}&=\frac{j_1+j_3}{5z}=\frac{1}{15}j_0+\frac{2}{21}j_2+\frac{1}{35}j_4\\
\end{split}
\end{equation}
which leads to:
\begin{equation}
\begin{split}
I^A_{1}(z)&=j_0(z)\\
I^A_{2}(z)&=-\frac{\sqrt{2}}{2}\,j_2(z)\\
I^A_{3}(z)&=\frac{1}{10}\,j_0(z)-\frac{1}{7}\,j_2(z)+\frac{9}{35}\,j_4(z)\\
I^A_{4}(z)&=\frac{1}{10}\,j_0(z)-\frac{1}{14}\,j_2(z)-\frac{6}{35}\,j_4(z)\\
I^A_{5}(z)&=\frac{1}{10}\,j_0(z)+\frac{1}{7}\,j_2(z)+\frac{3}{70}\,j_4(z)\\
\end{split}
\end{equation}
and
\begin{equation}
\begin{split}
I^B_{1}(z)&=0\\
I^B_{2}(z)&=0\\
I^B_{3}(z)&=\frac{2}{5}\,j_0(z)+\frac{4}{7}\,j_2(z)+\frac{6}{35}\,j_4(z)\\
I^B_{4}(z)&=\frac{2}{5}\,j_0(z)+\frac{2}{7}\,j_2(z)-\frac{4}{35}\,j_4(z)\\
I^B_{5}(z)&=\frac{2}{5}\,j_0(z)-\frac{4}{7}\,j_2(z)+\frac{1}{35}\,j_4(z)\\
\end{split}
\end{equation}

From these expressions, the real space RT correlation is obtained using~(\ref{eq:RT:all}). It is convenient to introduce the following transforms:
\begin{equation}\label{def:Cm}
\begin{split}
\mathring{C}^{(m)}(r)&=\frac{1}{2\pi^2}\int_0^\infty\d k\ k^2\,\mathring{\widehat{C}}(k)\,j_m(kr)
\end{split}
\end{equation}
We thus finally obtain the following expressions:
\begin{equation}
\begin{split}
\mathring{{C}}_{1}(r)&=\mathring{C}_1^{(0)}\\
\mathring{{C}}_{2}(r)&=-\frac{\sqrt{2}}{2}\,\mathring{C}_1^{(2)}\\
\mathring{{C}}_{3}(r)&=\frac{\mathring{C}_1^{(0)}+4\mathring{C}_5^{(0)}}{10}-\frac{\mathring{C}_1^{(2)}-4\mathring{C}_5^{(2)}}{7}+\frac{9\mathring{C}_1^{(4)}+6\mathring{C}_5^{(4)}}{35}\\
\mathring{{C}}_{4}(r)&=\frac{\mathring{C}_1^{(0)}+4\mathring{C}_5^{(0)}}{10}-\frac{\mathring{C}_1^{(2)}-4\mathring{C}_5^{(2)}}{14}-\frac{6\mathring{C}_1^{(4)}+4\mathring{C}_5^{(4)}}{35}\\
\mathring{{C}}_{5}(r)&=\frac{\mathring{C}_1^{(0)}+4\mathring{C}_5^{(0)}}{10}+\frac{\mathring{C}_1^{(2)}-4\mathring{C}_5^{(2)}}{7}+\frac{3\mathring{C}_1^{(4)}+2\mathring{C}_5^{(4)}}{70}
\end{split}
\end{equation}
which express the real-space correlation function in terms of functional transforms of $\mathring{\widehat{C}}_1(k)$ and $\mathring{\widehat{C}}_5(k)$. For the sake of completeness, let us just note that, using Eq.~(\ref{eq:C:0:inverse}), the above equations may be rewritten:
\begin{equation}
\begin{split}
\mathring{{C}}_{1}(r)&=C_0^{(0)}\\
\mathring{{C}}_{2}(r)&=-\frac{\sqrt{2}}{2}\,C_0^{(2)}\\
\mathring{{C}}_{3}(r)&=C_0^{\prime\,(0)}-\frac{2C_0^{(2)}-10C_0^{\prime\,(2)}}{7}+\frac{3C_0^{(4)}+6C_0^{\prime\,(4)}}{14}\\
\mathring{{C}}_{4}(r)&=C_0^{\prime\,(0)}-\frac{C_0^{(2)}-5C_0^{\prime\,(2)}}{7}-\frac{C_0^{(4)}+2C_0^{\prime\,(4)}}{7}\\
\mathring{{C}}_{5}(r)&=C_0^{\prime\,(0)}+\frac{2C_0^{(2)}-10C_0^{\prime\,(2)}}{7}+\frac{C_0^{(4)}+2C_0^{\prime\,(4)}}{28}
\end{split}
\end{equation}
which now expresses the correlation function in terms of transforms of the functions $\widehat{C}_0$ and $\widehat{C}_0'$. Of course $C_0^{(0)}$ and $C_0^{\prime\,(0)}$ are just $C_0$ and $C_0'$ (resp.).

\end{document}